\documentclass[11pt]{article}
\usepackage[utf8]{inputenc}
\usepackage[T1]{fontenc}
\usepackage{graphicx}
\usepackage{dcolumn}
\usepackage{bm}
\usepackage{chngpage}
\usepackage{braket}
\usepackage[margin=0.9in]{geometry}
\usepackage{amsmath}
\usepackage{soul}
\usepackage [english]{babel}
\usepackage [autostyle, english = american]{csquotes}
\MakeOuterQuote{"}

\usepackage{float}
\restylefloat{table}
\usepackage{placeins}
\usepackage{caption}
\usepackage{amsmath, amssymb}
\usepackage{bbold}
\usepackage{makecell}
\usepackage{braket}
\usepackage{calc}
\usepackage{subcaption}
\usepackage{authblk}
\usepackage{natbib}
\usepackage{indentfirst}
\usepackage[labelfont=bf]{caption}
\usepackage[dvipsnames]{xcolor}
\usepackage{hyperref}
\hypersetup{
    colorlinks,
    citecolor=blue,
    filecolor=black,
    linkcolor=blue,
    urlcolor=black
}

\DeclareUnicodeCharacter{2009}{\,} 
\usepackage{tabularx, booktabs}
\usepackage{multirow}
\usepackage{tikz}
\usepackage[left]{lineno}

\usepackage{geometry}
\usepackage{lastpage}
\usepackage{fancyhdr}
\newcolumntype{m}{>{\hsize=.24\hsize \centering\arraybackslash\arraybackslash}X}
\newcolumntype{n}{>{\hsize=.18\hsize \centering\arraybackslash\arraybackslash}X}
\newcolumntype{l}{>{\hsize=.18\hsize}X}
\newcolumntype{o}{>{\hsize=.15\hsize \centering\arraybackslash\arraybackslash}X}

\title{\textbf{Terahertz field-induced metastable magnetization near criticality in FePS$_3$}}

\author[1,*]{Batyr Ilyas}
\author[1,*]{Tianchuang Luo}
\author[1,*]{Alexander von Hoegen}
\author[2, 3,*]{Emil Vi\~nas Bostr\"om}
\author[4]{Zhuquan Zhang}
\author[5]{Jaena Park}
\author[5]{Junghyun Kim}
\author[5]{Je-Geun Park}
\author[4]{Keith A. Nelson}
\author[2, 6]{Angel Rubio}
\author[1,$\dag$]{Nuh Gedik}

\affil[1]{Department of Physics, Massachusetts Institute of Technology, Cambridge, 02139, Massachusetts, USA.}
\affil[2]{Max Planck Institute for the Structure and Dynamics of Matter, Luruper Chaussee 149, 22761 Hamburg, Germany.}
\affil[3]{Nano-Bio Spectroscopy Group, Departamento de Fisica de Materiales, Universidad del Pais Vasco, 20018 San Sebastian, Spain.}
\affil[4]{Department of Chemistry, Massachusetts Institute of Technology, Cambridge, 02139, Massachusetts, USA.}
\affil[5]{Department of Physics and Astronomy and Institute of Applied Physics, Seoul National University, Seoul 08826, Republic of Korea.}
\affil[6]{Center for Computational Quantum Physics, The Flatiron Institute, New York, NY 10010, USA.}
\affil[$\dag$]{e-mail: \texttt{gedik@mit.edu}}
\affil[*]{These authors contributed equally to this work.}

\begin{document}
\maketitle

\section*{Abstract}

Controlling the functional properties of quantum materials with light has emerged as a frontier of condensed matter physics, leading to discoveries of various light-induced phases of matter, such as superconductivity \cite{Mitrano2016PossibleTemperature}, ferroelectricity \cite{Nova2019MetastableSrTiO3, Li2019TerahertzSrTiO3}, magnetism \cite{McLeod2020Multi-messengerManganite, Disa2020PolarizingField, Disa2023Photo-inducedYTiO3} and charge density waves \cite{Kogar2020Light-inducedLaTe3}. However, in most cases, the photoinduced phases return to equilibrium on ultrafast timescales after the light is turned off, limiting their practical applications. In this study, we use intense terahertz pulses to induce a metastable magnetization with a remarkably long lifetime of over 2.5 milliseconds in a van der Waals antiferromagnet, FePS$_3$. The metastable state becomes increasingly robust as the temperature approaches the antiferromagnetic transition point, suggesting a significant role played by critical order parameter fluctuations in facilitating extended lifetimes. By combining first principles calculations with classical Monte Carlo and spin dynamics simulations, we find that the displacement of a specific phonon mode modulates the exchange couplings in a manner that favors a ground state with finite magnetization close to the N\'eel temperature. This analysis also clarifies how the critical fluctuations of the dominant antiferromagnetic order can amplify the magnitude and the lifetime of the new magnetic state. Our discovery demonstrates the efficient manipulation of the magnetic ground state in layered magnets through non-thermal pathways using terahertz light, and establishes the regions near critical points with enhanced order parameter fluctuations as promising areas to search for metastable hidden quantum states.

\section*{Introduction}
Thermal, quantum and order parameter fluctuations all play a critical role in shaping the phase diagrams of quantum materials. They typically arise close to phase transitions where the relative energy scales of different microscopic degrees of freedom become comparable. These strong fluctuations can hinder or facilitate the emergence of long-range order. For example, the antiferromagnetic fluctuations in underdoped cuprates are suggested to be the origin of the enigmatic pseudogap phase \cite{Schmalian1998WeakSuperconductors, Sedrakyan2010PseudogapFluctuations, Ye2019HubbardFluctuations}. On the other hand, strong quantum fluctuations of the atomic positions in SrTiO$_3$ prevent a long-range ferroelectric ordering at any temperature \cite{Miiller1991IndicationSrTi03}. Fluctuations also offer the ability to actively control the material properties by applying external stimuli such as pressure, strain, electric fields, or by tailoring the materials dielectric environment~\cite{Latini2021,Bostrom2022}, which can tune the relative energy scales of the system in favor of a new ground state. 

In addition to these static tuning knobs, photo-excitation by intense light pulses has emerged as a novel dynamical method of control, resulting in numerous light-induced symmetry broken states of matter. However, the lifetimes of these phases are predominantly in the picoseconds to nanoseconds range \cite{Mitrano2016PossibleTemperature, Li2019TerahertzSrTiO3, Disa2020PolarizingField, Kogar2020Light-inducedLaTe3, Afanasiev2021UltrafastPhonons}, with few examples of metastable cases \cite{Nova2019MetastableSrTiO3, McLeod2020Multi-messengerManganite, Stojchevska2014UltrafastCrystal, Disa2021OpticalYTiO3}. Finding viable strategies for stabilizing non-equilibrium states remains an ongoing and complex task. Moreover, high photon energy laser pulses, used in most of the experiments, deposit significant amount of thermal energy and perturb the underlying order, impeding deterministic control \cite{delaTorre2021Colloquium:Materials}. In this respect, terahertz pulses  have recently garnered  wide attention. The low photon energy of these sources enable selective excitation of collective modes to large amplitudes while keeping the orbital and electronic degrees of freedom in the ground state.

In this work, we demonstrate a light-induced metastable magnetization in the layered antiferromagnetic insulator FePS$_3$, with a lifetime of 2.5 milliseconds. We achieve this by resonantly driving its low-energy magnon and phonon modes to high amplitudes through illumination with intense terahertz (THz) pulses. This in turn modifies the exchange parameters driving the system into a state with a finite magnetization. The THz photon energy ($<$ 25 meV) is orders of magnitude smaller than the electronic gap of FePS$_3$ ($\sim$1.6 eV) \cite{Brec1979PhysicalChalcogenophosphites} and therefore preserves the low-energy physics and strong spin-lattice coupling. We find that the lifetime and amplitude of the terahertz-light induced magnetization diverges close to the N\'eel temperature, as a result of its coupling to the static and dynamic critical fluctuations of the dominant antiferromagnetic order parameter. Our results thereby highlight the crucial role that critical order parameter fluctuations play in stabilizing this non-equilibrium state.

\section*{Results}

\begin{center}
\begin{figure}
   \sbox0{\includegraphics[width=\textwidth]{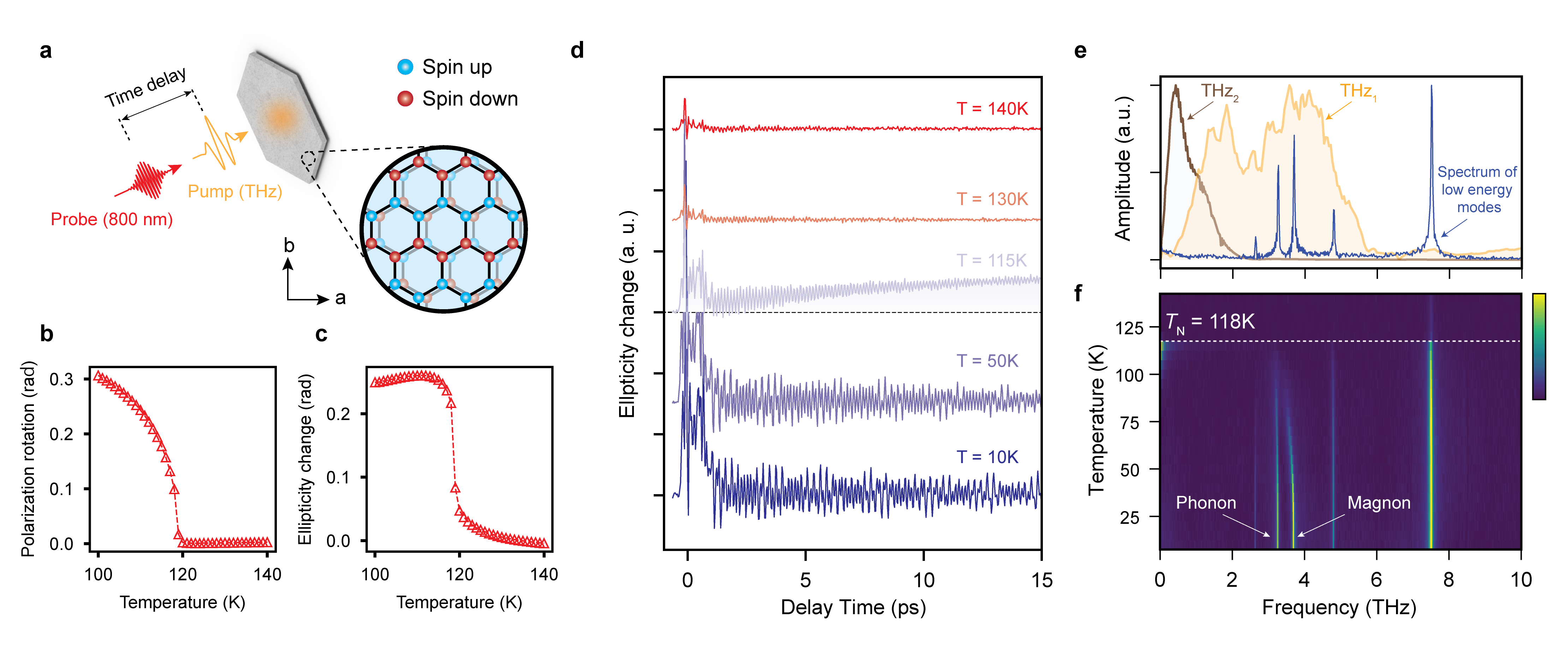}}
    \begin{minipage}{\wd0}
  \usebox0
  \captionsetup{labelfont={bf},labelformat={default},labelsep=period,name={Fig.}}
  \caption{\textbf{Experimental schematics and THz field-driven low energy modes.} \textbf{a.} Intense THz pulse (orange) drives low energy collective excitations in $\mathrm{FePS_3}$ and the induced transient changes in optical properties are probed with 800 nm probe pulse (red). $\mathrm{Fe^{2+}}$ ions form a hexagonal lattice and their spins arrange ferromagnetically along the zig-zag chain (\textit{a}-axis) and antiferromagnetically between the adjacent chains (the structure in the circle). The magnetic coupling between the layers is antiferromagnetic together with a small interlayer shear distortion along the $a$-axis. \textbf{b.} and \textbf{c.} are temperature dependencies of polarization rotation and ellipticity change signals in equilibrium, respectively. \textbf{d.} Probe ellipticity change signal traces at selected temperatures exhibit coherent oscillations induced by THz-field. As temperature approaches the transition point (e.g. at T = 115K), the time traces start to build a net-positive response. \textbf{e.} Fourier transform of the oscillations at T = 10K (dark blue) contains 5 prominent peaks that correspond to a magnon and phonon modes. The spectral content of the excitation pulse is shown in orange, labeled as THz$_1$. For additional experiments a THz field generated from a different source is used, with a spectrum shown in brown and labeled as THz$_2$, and it is not resonant with any of the modes.  \textbf{f.} Temperature evolution of the Fourier spectrum. Colorbar is normalized. A notable change in the spectra occurs below the N\'eel temperature ($T\mathrm{_N}$ = 118 K) and the magnon mode softens upon increasing temperature.
  \label{fig:expsetup}}
\end{minipage}
\end{figure}  
\end{center}

\textit{M}PS$_3$ (where \textit{M} = Fe, Ni, or Mn) is a group of honeycomb lattice van der Waals antiferromagnetic (AFM) materials, which host stable AFM orders down to atomically thin limits \cite{Lee2016Ising-TypeFePS3, Kim2019SuppressionNiPS3, Kim2019AntiferromagneticSpectroscopy, Long2020PersistenceCrystals}. The specific choice of the transition metal \textit{M} determines the strength and the sign of the exchange interactions which gives rise to diverse types of spin orderings \cite{LeFlem1982, Joy1992, Sivadas2015}. The spin structure exhibits a complex interplay with other degrees of freedom such as the lattice and orbital ordering, which manifested as distinct fingerprints in recent experiments \cite{Lee2016Ising-TypeFePS3, Ergecen2023CoherentAntiferromagnet, Zhou2022DynamicalAntiferromagnets, Kang2020CoherentNiPS3, Hwangbo2021HighlyInsulator, Ergecen2022MagneticallyAntiferromagnet, Belvin2021Exciton-drivenInsulator, Afanasiev2020ControllingLight, Khusyainov2023UltrafastCoPS3b}. FePS$_3$ stands out among the other members of the \textit{M}PS$_3$ family in two characteristics. First, it features a pronounced single-ion anisotropy \cite{Wildes2020Evidence3, Wildes2012TheFePS3} arising from spin-orbit coupling enabled by unquenched orbital moments of Fe$^{+2}$ ions. Together with a ferromagnetic nearest neighbor and antiferromagnetic third nearest neighbor exchange interaction, this results in a zig-zag AFM ground state below $T_\textrm{N} = 118$ K. The large single-ion anisotropy forces the magnetic moments to align in the out of plane direction, which then arrange ferromagnetically within the zig-zag chains (\textit{a}-axis), and antiferromagnetically between the adjacent chains (see Fig.~\ref{fig:expsetup}a, the structure in the circle). Second, FePS$_3$ exhibits a particularly strong spin-lattice interaction \cite{Lee2016Ising-TypeFePS3, Joy1992, Ergecen2023CoherentAntiferromagnet, Zong2023}, which is exemplified by recent magneto-Raman experiments that revealed strong hybridization between the low energy magnon and phonon modes \cite{mccreary2020, liuPRL2021, vaclavkova2021, Zhang2021CoherentInsulator}. This may offer an  effective pathway to manipulate the magnetic order by distorting the crystal lattice either statically by applying strain or pressure or dynamically by driving the relevant lattice modes to large amplitudes.

Recent experiments tried to leverage the dynamical approach to achieve an ultra-fast manipulation of the magnetism in FePS$_3$ through excitation of the electronic subsystem with high photon energy light pulses  \cite{Ergecen2023CoherentAntiferromagnet, Zhou2022DynamicalAntiferromagnets, Mertens2023UltrafastFePS3, Zhang2021SpinFePS3, Zong2023}. This excitation couples only indirectly to the lattice, yet is still effective in driving coherent oscillations of the lattice (phonons) and spins (magnons) and demonstrated an ultra-fast weakening of the antiferromagnetic order. Here we use an alternative strategy to leverage the strong spin-lattice coupling. We drive the hybridized spin and lattice modes directly with intense THz pulses. To study the role of resonant driving, we use two THz sources generated via optical rectification in different nonlinear crystals (see Methods). For resonant excitation, we use a THz field with a broad spectrum that covers the range of 0.2 - 6 THz (Fig.~\ref{fig:expsetup}e, shown in orange, labeled as "THz$_1$"), with a peak field of $\sim 300$ kV/cm (Extended Data Fig.~\ref{fig:expsetupscheme}b). For off-resonant excitation, we use a THz source with a spectrum primarily concentrated at energies below $\sim 2$ THz (Fig.~\ref{fig:expsetup}e, shown in brown, labeled as "THz$_2$"), which is not resonant with any of the lattice or spin excitations. The peak field in this case is $\sim 600$ kV/cm. We monitor the ensuing dynamics induced by the THz pump pulse by using a weak near-infrared (800 nm) probe pulse with variable time delay $\Delta t$ and analyzing the polarization state of the transmitted beam (see Fig. \ref{fig:expsetup}a). Specifically, we measure the ultra-fast polarization rotation and ellipticity of the initially linearly polarized probe pulses.

We establish the relationship between these polarization characteristics and the underlying magnetic order by measuring the temperature dependence of their equilibrium values. The static polarization rotation signal, $\theta_\textrm{st.}(T)$, sets on below the transition temperature $T_{\rm N}$ and has an order parameter-like behavior (Fig.~\ref{fig:expsetup}b), indicating its sensitivity to the AFM order. The static ellipticity change of the near-infrared light, $\eta_\textrm{st.}(T)$, on the other hand, has a markedly different temperature dependence. Below the N\'eel temperature, $\eta_\textrm{st.}(T)$ is strongly enhanced, but it also shows a finite tail above $T_{\textrm{N}}$ which is absent in the polarization rotation signal, suggestive of sensitivity to magnetic fluctuations (Fig.~\ref{fig:expsetup}c). Quantitatively, the ellipticity change and polarization rotation measure the anisotropy in the real and imaginary parts of the complex refractive index, respectively (see Supplementary Notes 11 and 12). Therefore, by measuring these two quantities, we can comprehensively study the THz-induced magnetic dynamics in FePS$_3$. We note that similar measurements in equilibrium were performed on other systems \cite{Wang2021Spin-inducedCrystals, Khusyainov2023UltrafastCoPS3b, Zhang2021ObservationFePS3}, and have established these optical anisotropy probes as a good proxy of the zig-zag AFM order.

Fig.~\ref{fig:expsetup}d shows representative traces of the time-resolved ellipticity change at selected temperatures. The transient changes induced by the  broadband THz pulse ("THz$_1$") consist of fast oscillations and a strong positive signal near time zero. The positive signal near zero time delay, with a positive background, occurs as a result of the so called terahertz Kerr effect and marks the temporal overlap of pump and probe pulses \cite{Hoffmann2009TerahertzEffect}. This is followed by oscillatory features at later times. The Fourier transform of these oscillations at $10$ K is shown in Fig.~\ref{fig:expsetup}e (blue). We can identify four distinct phonon modes and a magnon, the frequencies of which are in good agreement with Raman \cite{liuPRL2021} and infrared spectroscopies \cite{Zhang2021CoherentInsulator} (see Extended Data Table~\ref{fig:RamanIRTable}). Most of these modes are spectrally covered by our broad THz excitation spectrum (Fig.~\ref{fig:expsetup}e in orange). We note that several phonons observed in our experiments are found to be visible both in Raman and infrared spectra, despite the fact that FePS$_3$ preserves inversion symmetry \cite{Lancon2016c}. This is explained by the strong hybridization of the magnon and phonon modes, which alters the selection rules \cite{Zhang2021CoherentInsulator} and allows these modes to be visible in both of these experimental probes.  The defining features of the transient ellipticity signal at low temperatures are in qualitative agreement with the polarization rotation signal (see Extended Data Fig.~\ref{fig:modesHWPQWP}).

\begin{center}
\begin{figure}
   \sbox0{\includegraphics[width=\textwidth]{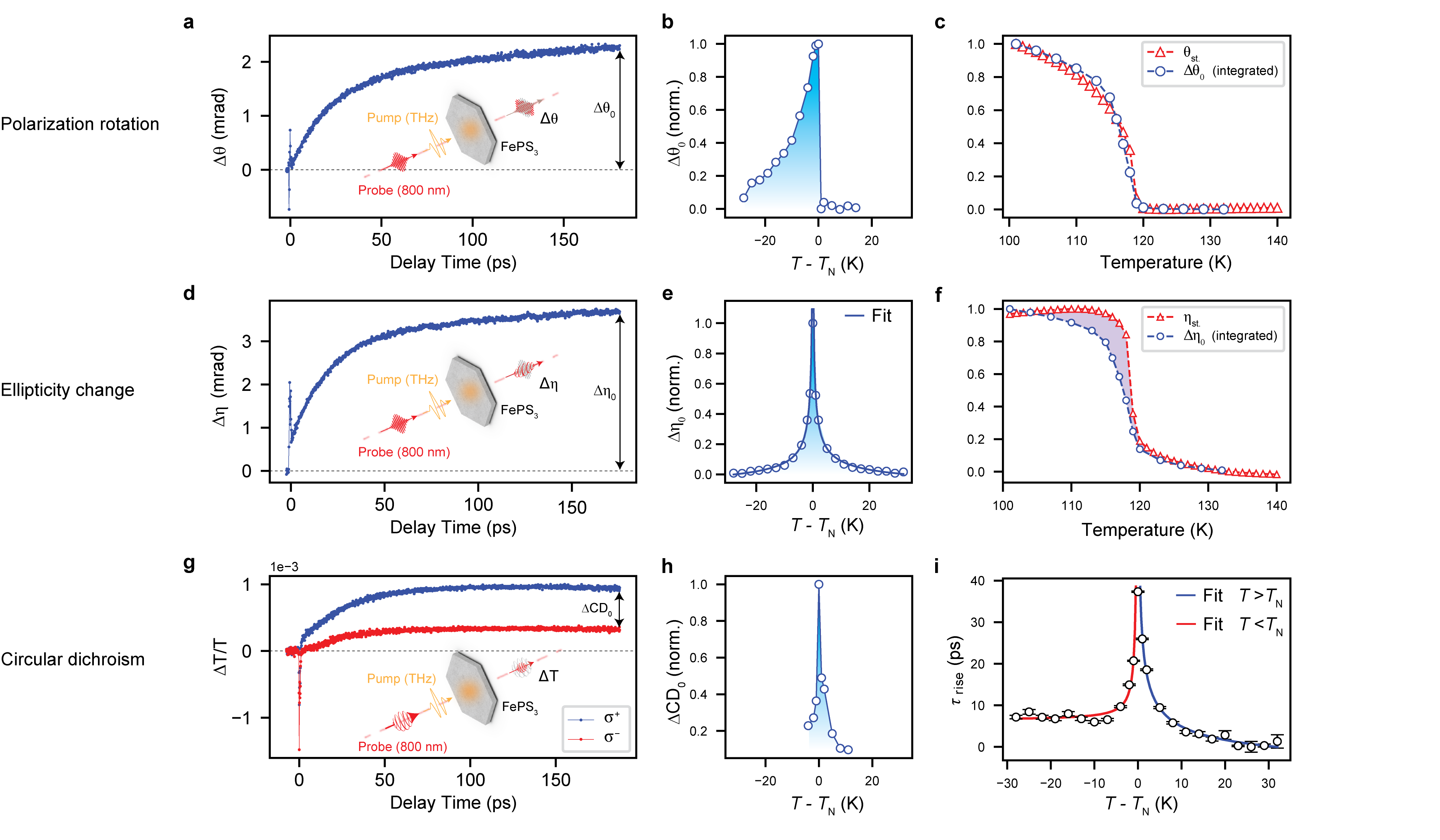}}
    \begin{minipage}{\wd0}
  \usebox0
  \captionsetup{labelfont={bf},labelformat={default},labelsep=period,name={Fig.}}
  \caption{\textbf{THz field-induced nonequilibrium state with a net magnetization.}  THz field induced long-lived polarization rotation \textbf{a}, ellipticity change \textbf{d}, and circular dichroism \textbf{g} signals, all measured at 118K. Note: all three time traces are shifted vertically, so that the pre-time zero signal is at zero level. The temperature dependent raw data are provided in Extended Data Fig.~\ref{fig:rawtimetrace}. The temperature dependence of the maximum signal values labeled as $\Delta \theta_0(T)$, $\Delta \eta_0(T)$, and $\Delta$CD$_0(T)$, measured at $\sim$170 ps, are given in \textbf{b}, \textbf{e}, and \textbf{h}, respectively. To analyze the THz induced heating effect, we measured static polarization rotation and ellipticity change signals, as shown in \textbf{c} and \textbf{f} (red). Further, we integrate the dynamic changes in of these optical properties (\textbf{b}, \textbf{e} and \textbf{h}) in temperature and compare with static values. In the case of polarization rotation the static and dynamic responses (integrated) coincide well, indicating THz induced response is mostly due to heating effect. In the case of ellipticity change, there is a sizeable contrast between the two responses, hinting at an additional nonthermal effect. This observation together with $\Delta$CD response in \textbf{h}, imply a nonequilibrium state with a net magnetization. The $\Delta \eta_0(T)$ in \textbf{i}, is fit by a power law (see Methods) and the exponent is 0.56 $\pm$ 0.05. \textbf{i.} The critical behavior of the rise time dynamics is fit with a power law (Eq.~\ref{eq:tauFit}), and the obtained critical exponents are $\nu z$ = 1.07 $\pm$ 0.27 and $\nu z$ = 0.44 $\pm$ 0.07, for temperatures below and above $T_\textrm{N}$, respectively.  
  \label{fig:faradayCD}}
\end{minipage}
\end{figure}  
\end{center}

Next, we turn to the temperature dependence of the coherent modes (see Fig.~\ref{fig:expsetup}f) and the dynamics close to the N\'eel temperature. The spectrum of the coherent oscillations undergoes an abrupt change below the N\'eel temperature and the magnon mode softens upon temperature increase. In particular, the observed phonon modes with energies below $\sim 5$ THz are found to vanish above $T_{\rm N}$, indicating that they are zone-folded modes induced by the zigzag AFM order. Notably, as we approach the transition temperature, a long-lived signal starts to build up in response to the THz pump, as can be seen in the 115 K trace in Fig.~\ref{fig:expsetup}d. This signal manifests as a low frequency spectral weight near  $T_{\rm N}$ in Fig.~\ref{fig:expsetup}f. To further analyze the nature of this long-lived state, we extend the delay time window up to a maximum of $\sim175$ ps and observe that at $T_{\textrm{N}} = 118$ K the signal is still growing with no sign of saturation, both in the polarization rotation $\Delta \theta (t)$ (Fig.~\ref{fig:faradayCD}a) and the ellipticity change $\Delta \eta (t)$ (Fig.~\ref{fig:faradayCD}d) channels. Fig.~\ref{fig:faradayCD}b and e show the maximum values of these quantities extracted at $t = 175$ ps, $\Delta \theta_0(T)$ and $\Delta \eta_0(T)$, as a function of temperature. Both quantities peak at the N\'eel temperature, however they display distinct behaviors: $\Delta \theta_0(T)$ has an asymmetric shape across $T_\textrm{N}$, whereas $\Delta \eta_0(T)$ is symmetric. 

To identify the degrees of freedom that are responsible for the long-lived signal, we first consider the possibility of electronic excitations. In our experimental setting, there are two potential pathways of carrier excitation: (i) multiphoton absorption and (ii) tunneling. The multiphoton absorption process is highly suppressed due to the $2 - 3$ orders of magnitude mismatch in energies of the gap ($\sim 1.6$ eV) \cite{Brec1979PhysicalChalcogenophosphites} and THz photons ($< 25$ meV). To explore the tunneling scenario, which is more efficient at higher electric fields, we repeat the experiments with more intense single-cycle THz pulses which spectrally do not overlap with any of the low energy modes (see Fig.~\ref{fig:expsetup}e, "THz$_2$"). The induced dynamics did not show any long-lived state near $T_{\textrm{N}}$ (Extended Data Fig.~\ref{fig:LNOdata}b). Therefore we rule out the possibility of electronic excitations, which in turn indicates that the long-lived signal must be triggered by resonant excitation of the low energy modes. This observation highlights the fundamental difference between the THz drive and the previous works with above-gap photo-excitations \cite{Zhou2022DynamicalAntiferromagnets, Zhang2021SpinFePS3}.

Since the THz spectrum covers several mode resonances, the temperature increase in the lattice or in the spin subsystems, and consequently the reduction of the order parameter cannot be ignored. To understand the role of this heating effect, we compare the THz-induced dynamics and the static optical responses. For this purpose, we integrate the temperature dependent functions $\Delta \theta_0(T)$ and $\Delta \eta_0(T)$ (Fig.~\ref{fig:faradayCD}b and e) from a high temperature cut-off $T_{\rm high}$ down to the temperature $T$, and plot the resulting functions $\theta_{\rm int}(T) = \int_T^{T_{\rm high}} \Delta \theta_0(T)$ and $\eta_{\rm int}(T) = \int_T^{T_{\rm high}} \Delta \eta_0(T)$ together with the static values $\theta_{\rm st.}(T)$ and $\eta_{\rm st.}(T)$. In the polarization rotation channel, the static and the integrated dynamic responses are nearly identical to each other (Fig.~\ref{fig:faradayCD}c)  indicating that the change $\Delta \theta_0(T)$ is mainly due to a THz-induced change in temperature. Since $\theta_{\textrm{st.}} (T)$ is sensitive to the AFM order parameter, $\Delta \theta_0 (T)$ can be assigned to a thermal reduction of the AFM order parameter. Additionally, lattice distortions due to nonlinearly excited phonons can also reduce the AFM order and lead to a polarization rotation signal. On the other hand the behavior in the ellipticity channel is distinct (Fig.~\ref{fig:faradayCD}f). There is a considerable deviation of the THz-induced response from the static signal, which implies the presence of an additional non-thermal contribution. Notably, the sharp divergences of $\Delta \theta_0(T)$ and $\Delta \eta_0(T)$ near $T_{\textrm{N}}$ indicate an infinitesimal increase in the spin temperature induced by the THz field, since a significant heat deposition would lead to a broad peak shifted away from $T_{\rm N}$, as observed in works with above gap excitation \cite{Zhang2021SpinFePS3}.

The ellipticity of our initially linearly polarized light can be caused by two distinct optical mechanisms, linear birefringence and circular dichroism (CD). However, in a fully-compensated antiferromagnet with zero net magnetization, such as FePS$_3$, we do not expect an equilibrium CD in the optical range. Moreover, a CD response can only occur in certain nonmagnetic media, such as in systems with a chiral lattice structure.  However, detailed characterizations of the crystal structure of FePS$_3$ \cite{Zhou2022DynamicalAntiferromagnets, Ouvrard1985StructuralCd} show no sign of chiral behavior. To investigate the possibility of a nonequilibrium CD, we study the response of circularly polarized probe pulses to THz excitation. We find a significant difference in the relative transmission of left- ($\sigma_-$) and right-handed ($\sigma_+$) circularly polarized light (Fig.~\ref{fig:faradayCD}g), i.e. a circular dichroism, with a slow rise-time similar to the transient ellipticity. Additionally, the temperature dependence of the transient CD (Fig.~\ref{fig:faradayCD}h) resembles that of the ellipticity, suggesting that the ellipticity signal is indeed caused by a THz-induced CD. The observed out-of-equilibrium dichroism signal, $\Delta$CD$(T)$, therefore indicates that the new photoinduced state has a net out-of-plane magnetization, which coexists with the weakly suppressed but still dominant zig-zag AFM order. 

\begin{center}
\begin{figure}
   \sbox0{\includegraphics[width=0.6\textwidth]{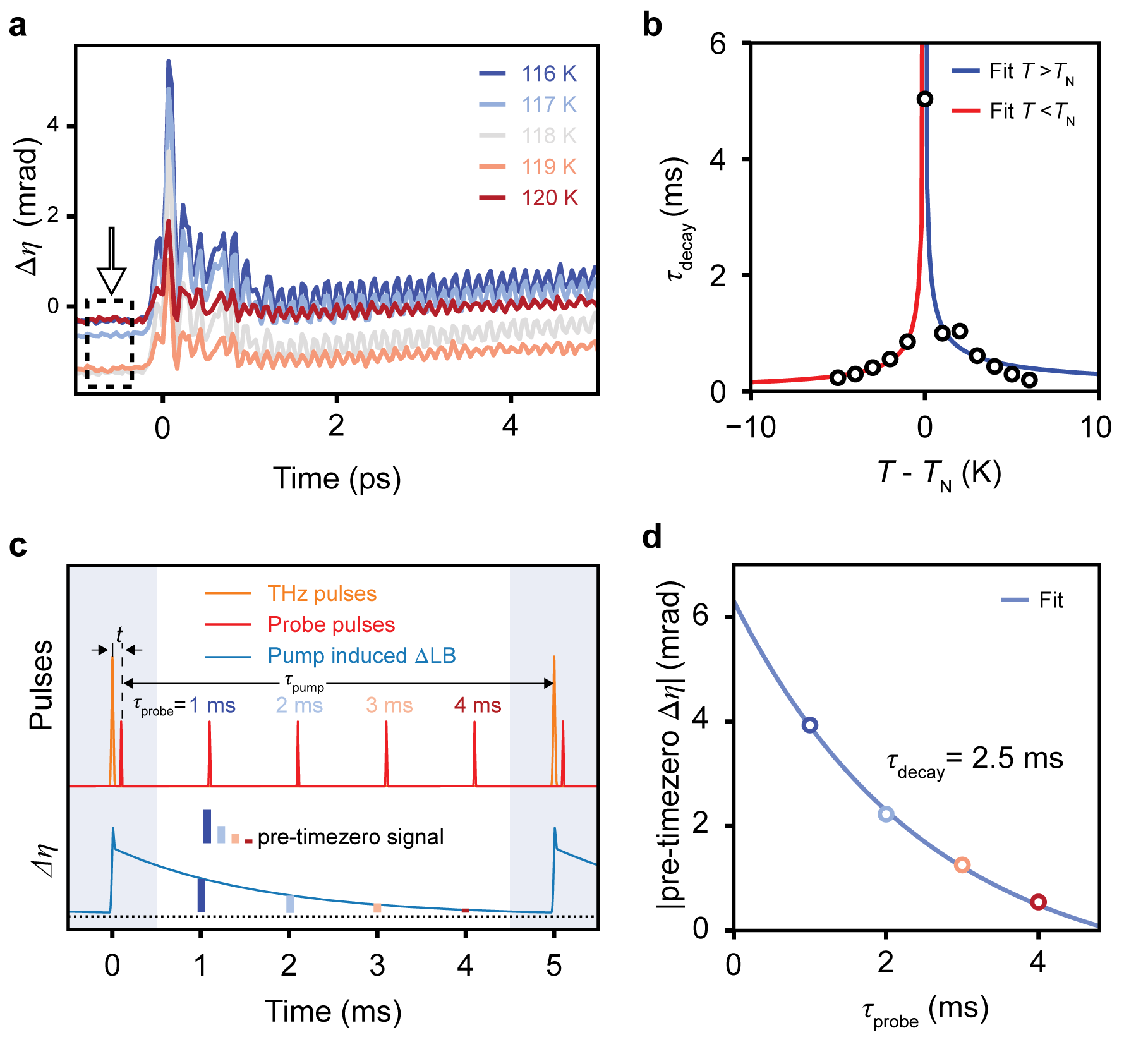}}
    \begin{minipage}{\wd0}
  \usebox0
  \captionsetup{labelfont={bf},labelformat={default},labelsep=period,name={Fig.}}
  \caption{\textbf{Decay time of the photoinduced state  exceeds 2 ms.} \textbf{a}. Pre-time zero signal (dashed square under the arrow) around $T_\mathrm{N}$ in the ellipticity channel.\textbf{b.} Metastable state decay time obtained from the pre-time zero signal. The red and blue lines are power law fits which gives $z\nu=0.89$ and $0.56$. \textbf{c}. Studying the decay dynamics with a different THz repetition rate. The upper panel shows the train of THz pump (orange) and optical probe (red) pulses. The pump pulses are separated by $\tau_\text{pump}=5~\text{ms}$. The probe pulse separation are chosen between  $\tau_\text{probe}=$ 1 ms, 2 ms, 3 ms, and 4 ms. $t$ labels pump-probe time delay. The experimentally accessible region of $t$ is much shorter than pulse-to-pulse separations. The blue shaded region is the pump-on region, and the unshaded region is pump-off. The lower panel shows the pump induced $\Delta \eta$. The cumulative effect of the pump pulses leads to a net offset from zero. The colored bars label the strength of the pre-time zero signal in this measurement configuration, obtained by the difference between pump-on (shaded) and pump-off (unshaded) region. \textbf{d}. The fitting of pre-time zero signal as a function of $\tau_\text{probe}$, which gives an estimation of the intrinsic relaxation time $\tau_{\textrm{decay}}$ of the metastable state. Colored markers are the absolute values of the pre-time zero signal in \textbf{c}. 
  \label{fig:dutycycle}}
\end{minipage}
\end{figure}  
\end{center}

Next we examine the dynamics of the induced nonequilibrium magnetic state. We note that in Fig.~\ref{fig:faradayCD}a, d, and g, the time traces are shifted vertically, such that all the traces are set to zero before time zero. However, in the raw data shown in Fig.~\ref{fig:dutycycle}a and Extended Data Fig.~\ref{fig:rawtimetrace}, as we approach $T_\textrm{N}$, a significant amount of pre-time zero signal develops. The existence of a pre-time zero signal hints at a relaxation time $\tau_{\mathrm{decay}}$ comparable with the separation time between the probe and pump pulses (1 ms and 2 ms). In this case, the effects of consecutive pump pulses accumulate and lead to an ellipticity change of the probe even in the pump-off region. A larger ratio between the pre-time zero signal and the maximal signal after time zero indicates a slower decay of the pump induced $\Delta\eta$. From the modeling detailed in the Methods section, $\tau_\mathrm{decay}$ can be extracted, and its temperature evolution is shown in Fig. \ref{fig:dutycycle}b. A clear divergence around $118$ K is observed. However, we note that the value of $\tau_\mathrm{decay}$ extracted here is an overestimation since the maximal value of the signal after time zero is not reached in the measurement time window (30 ps). To further substantiate our model and obtain a more accurate $\tau_\mathrm{decay}$, we extend the temporal separation between the THz pulses ($\tau_{\mathrm{pump}}$) from 2 ms to 5 ms (the orange trace in Fig.~\ref{fig:dutycycle}c), allowing the system to fully relax to equilibrium conditions before the arrival of the subsequent THz pulse. As we vary the time of the probe in the pump-off region ($\tau_\mathrm{probe}$) from 1 ms to 4 ms, the pre-time zero signal gradually decays (Fig.~\ref{fig:dutycycle}d), corresponding to the decay of $\Delta\eta$ in the pump-off region (the blue trace in Fig.~\ref{fig:dutycycle}c). By fitting the pre-time zero signal for different $\tau_\mathrm{probe}$ with the same model (see Methods), we find the relaxation time of the metastable state at $118$ K to be $\tau_{\mathrm{decay}} \sim 2.5$ ms. Therefore the THz-induced state with a net magnetization has a remarkably long lifetime.

The comparison between a resonant and off-resonant excitation of the low energy modes (Extended Data Fig.~\ref{fig:LNOdata}), as performed above, highlights the importance of large amplitude displacements of these modes in inducing the long-lived magnetic state. In contrast, linear excitation of infrared-active phonons results in an oscillatory motion of the atoms around their equilibrium positions, and cannot give rise to the accumulation of net positive signal over long delay times ($\sim 100$ ps). A net positive or negative signal instead requires a net average distortion of the lattice, which can be realized through various nonlinear excitation pathways \cite{Juraschek2018Sum-frequencyScattering}. To identify the modes that are nonlinearly driven, we analyze the field strength dependence of mode amplitudes. The lowest energy phonon mode ($\Omega_1 = 2.64$ THz), as well as the magnon ($\Omega_m = 3.69$ THz), exhibit a linear field dependence, whereas the $\Omega_3 = 4.80$ THz and $\Omega_4 = 7.51$ THz phonons show  a quadratic dependence on the field (see Extended Data Fig.~\ref{fig:otherphononfluence}). The linear dependence of the magnon indicates a direct excitation by the magnetic component of the THz field. Remarkably, the field strength dependence of the phonon mode at $\Omega_2=$ 3.27 THz cannot be fit by either a linear or quadratic function, but instead displays a superposition of linear and quadratic behavior (see Fig.~\ref{fig:criticality}a and Supplementary Note 1). Therefore the nonlinearly excited phonon modes $\Omega_2$, $\Omega_3$, and $\Omega_4$ are the main candidates to explain the induced magnetization.


To better understand how nonlinear driving of these low-energy modes leads to metastable state with finite magnetization, we formulate a microscopic theory of the spin and lattice degrees freedoms in FePS$_3$. The magnetic structure is governed by the Hamiltonian~\cite{Zhang2021CoherentInsulator,Cui2023}
\begin{equation}\label{eq:spin_ham}
    H = \sum_{ij} J_{ij} \mathbf{S}_i \cdot \mathbf{S}_j - \Delta \sum_{i} S_{iz}^2,
\end{equation}
where $\mathbf{S}_i$ is the spin of the Fe ion at lattice site $i$, and $J_{ij}$ is the exchange coupling between spins at sites $i$ and $j$. The first term of $H$ represents the Heisenberg interaction, while the second term corresponds to an Ising-type single-ion anisotropy. When a phonon mode is excited this results in a change of the inter-ionic distances, which in turn modifies the exchange parameters. For small displacements the exchange interaction can be expanded as ${J}_{ij}(Q) \approx {J}_{ij} - \alpha_{ij} Q$, where $\alpha_{ij}$ quantifies the modulation of the magnetic interaction and $Q$ is the phonon displacement. The model is completely parameterized from first principles, and the calculated values of $J_{ij}$ and $\alpha_{ij}$ (see Extended Data Table \ref{tab:spin_phonon}) are found to correctly reproduce the equilibrium magnetic properties~\cite{Zhang2021CoherentInsulator} (see also Supplementary Note 6). In particular, from calculations within the frozen phonon approximation (see Methods) the spin-phonon couplings $\alpha_{ij}$ are found to decay rapidly with inter-atomic distance, and to be significant only for nearest neighbor spins.

The spatial patterns in which the phonons modulate the exchange parameters are shown in Fig.~\ref{fig:criticality}b and Extended Data Fig.~\ref{fig:otherphononfluence}. These spatial modulations are coarse grained to obtain a coupling between the phonon modes and the macroscopic magnetic variables $L = S_1 - S_2 - S_3 + S_4$ and $M = S_1 + S_2 + S_3 + S_4$, corresponding to the AFM and ferromagnetic (FM) order parameter, respectively (see also Supplementary Note 5.2). Here $S_i$ denotes the four spins of the magnetic unit cell, labeled according to Fig.~\ref{fig:criticality}b. From the coarse grained theory we find that only the $\Omega_2$ = 3.27 THz phonon mode can lead to a finite magnetization (see Supplementary Note 5 for a detailed analysis), via a coupling of the form $H_{\rm m-ph} = g LM Q_{2}$. The real space motions of the atoms corresponding to this mode modifies the bond lengths between the Fe atoms as shown in Fig.~\ref{fig:criticality}b, which enhances the exchange interaction within one of the FM chains while it weakens within the adjacent chain. This phonon induced magnetization, as well as the form of the coupling, is corroborated by Monte Carlo simulations of the microscopic spin Hamiltonian (Eq.~\ref{eq:spin_ham}), where the magnetic interactions are modulated according to the phonon displacement patterns (see Supplementary Note 6 and Fig. S2). We therefore conclude that the nonlinear driving of the $3.27$ THz phonon is responsible for the THz-induced magnetization observed here. Considering the spectrum of the THz pulse and the energies of linearly driven modes, the nonlinear excitation of 3.27 THz mode can happen via purely photonic or via infrared resonant Raman scattering processes \cite{Khalsa2021}. The purely ionic process is unlikely to take place due to energy matching constraints. On the other hand, since the magnetization is induced by the displacement of the 3.27 THz mode, this type of rectification process can also be mediated by infrared-active phonons \cite{Juraschek2018Sum-frequencyScattering}, as our two-dimensional terahertz spectroscopy studies (Fig. S5) revealed a coupling with 4.8 THz infrared-active phonon. This type of dynamical modulation of the exchange coupling parameters can be broadly applied to other magnetic materials exhibiting strong spin-phonon coupling \cite{Padma2021, Padma2022}.

\noindent An effective Ginzburg-Landau free energy can then be written as

\begin{align}\label{eq:ginzburg_landau}
 F &= \frac{a_L(T)}{2} L^2 + \frac{b_L}{4} L^4 + \frac{a_M(T)}{2} M^2 + \frac{b_M}{4} M^4 + \frac{\Omega^2}{2} Q_2^2 + g L M Q_2,
\end{align}
where only the effects of the $Q_2$ phonon are considered (see Supplementary Note 7). The free energy landscape as a function of $L$ and $M$, with and without a distortion along the $Q_2$ mode, is illustrated in Fig.~\ref{fig:criticality}c and d. The phonon distortion leads to a new displaced minimum with a net magnetization, $M \neq 0$ (red line), whose sign is determined by the sign of $Q_2L$. Therefore the application of an external magnetic field will not flip the sign of $M$, as opposed to common expectation for ferromagnets in equilibrium. Furthermore, the sign of the induced magnetization is insensitive to domain formation. This is attributed to the fact that the sign of lattice distortion, $Q_2$, is connected to the sign of $L$ (see Supplementary Note 13 for a detailed explanation), such that the product $Q_2 L$ is set by material parameters only.



As demonstrated below, this Ginzburg-Landau theory can explain the temperature dependence of both the magnitude and lifetime of the THz-induced magnetization. Minimizing the free energy with respect to $M$ and $Q_2$, assuming $L$ is constant, the global energy minimum is reached for nonzero values of both $M$ and $Q_2$. In particular, the magnetization of the global minimum is proportional to the variance of AFM order parameter, $M \sim \sqrt{\langle L^2\rangle}$ (see Supplementary Note 6). Since the variance of $L$ is proportional to the spin structure factor, which in turn is related to magnetic susceptibility of the zigzag order (see Supplementary Note 4), the magnetization is expected to scale with temperature as $M \sim  \sqrt{\chi_{zz}}$. As a result, the magnitude of the nonequilibrium magnetization is expected exhibit a critical divergence near $T_\textrm{N}$ \cite{Joy1992}, such that
\begin{align}
 M(T) \sim |T-T_{\textrm{N}}|^{-\gamma/2}.
\end{align}
Here $\gamma$ is the critical exponent of the magnetic susceptibility. Here we note that near the transition point, the critical fluctuations dominate the dynamics, and fluctuations of the $Q_2$ phonon are also present above $T_{\textrm{N}}$. To compare the response expected from the 3-D Ising universality class with the experimental values, we assume that $\Delta \eta_0(T)$ acts as a probe of the magnetization. This assumption is further corroborated by the similar temperature dependencies of $\Delta \eta_0(T)$ and $\Delta$CD$_0$, the latter of which is a direct measure of $M(T)$. The power-law fit to the temperature dependence (see Methods) of the observed $\Delta \eta_0(T)$ gives a critical exponent $\gamma/2$ = 0.56 $\pm$ 0.05 (see Fig.~\ref{fig:faradayCD}e), which is in good agreement with the expected value $\gamma/2$ = 0.62 \cite{Ferrenberg2018PushingModel}. 

To explain the long-lived nature of the induced magnetization, we note that the Ginzburg-Landau theory in fact predicts a degenerate pair of shallow free energy minima at $\pm(M,Q_2)$, as shown in Fig.~\ref{fig:criticality}e (upper panel). In thermal equilibrium, the order parameter will fluctuate between the two minima, preventing a build-up of net magnetization in equilibrium (see Supplementary Note 7). This is similar to how quantum fluctuations prevent ferroelectric ordering in SrTiO$_3$~\cite{Miiller1991IndicationSrTi03}. However, when the system is driven into an asymmetric configuration by the THz-induced finite phonon displacement close to transition point (Fig.~\ref{fig:criticality}e middle panel), the fluctuating order parameters $L$ and $M$ are stabilized and start to migrate, such that $M$ localizes in the free energy minimum selected by the sign of $Q_2 L$ (Fig.~\ref{fig:criticality}e lower panel). This mechanism is supported by extensive microscopic spin dynamics simulations, showing a build-up of magnetization in response to dynamical phonon displacement (see Supplementary Note 9).

The slow decay of the induced perturbations will eventually bring the system back to thermal equilibrium. However, close to a second order phase transition, the relaxation of perturbations in the dominant order parameter are expected to show a critical slowing down following the exponential form $\delta L(t) \sim e^{-t/\tau(T)}$, with a lifetime $\tau(T) \sim |T - T_{\rm N}|^{-\nu z}$ diverging as $T \to T_{\rm N}$~\cite{Hohenberg1977}. As discussed in the Methods (and in detail in Supplementary Note 10), both a solution of the stochastic Ginzburg-Landau equations and an analytic solution of the deterministic equations show that for $T \to T_{\rm N}$ the induced magnetization $M(t)$ quickly relaxes to the free energy minimum determined by $\delta L$, and then adiabatically follows the free energy minimum back to equilibrium. Crucially, the relaxation of the free energy minimum is determined by the dynamics of the antiferromagnetic order, and therefore exhibits a critical slowing down. At long times the magnetization is therefore given by $M(t) \approx g/\sqrt{\Omega b_M} \delta L(t) \sim \exp(-t/\tau)$, such that $M(t)$ inherits its critical behavior from $\delta L$. In particular, the magnetization shows a diverging relaxation time close to $T_{\rm N}$ with the same dynamical exponents as $\delta L$. Analyzing the temperature dependence of both the rise time (Fig.~\ref{fig:faradayCD}i) and decay time (Fig.~\ref{fig:dutycycle}b), the dynamical critical exponents in the region $T < T_{\textrm{N}}$ are close to the expected values from 3-D Ising model (see Methods).

\section*{Conclusions}
In summary, we discovered a hidden state with finite magnetization induced by an intense THz field. This nonequilibrium phase emerges as a result of the nonlinear driving of a specific phonon mode, leading to significant modifications in the in-plane distances between the magnetic ions. We find that the equilibrium and dynamical critical fluctuations of the underlying antiferromagnetic order respectively enhance the magnitude of the nonequilibrium magnetization and facilitate its metastability. Remarkably, our theoretical analysis captures all aspects of the experimental observations. Notably the magnetization mechanism in this case is distinct from the dynamic piezomagnetic effect \cite{Disa2020PolarizingField, Formisano2022Laser-inducedCoF2}, as the magnetic point group of FePS$_3$ ($2/m1'$ \cite{Lancon2016c}) prohibits piezomagnetism. All these findings present an interesting new angle of phase transitions by showing that the regions near critical points in the phase diagram, with enhanced order parameter fluctuations, are promising areas to search for metastable hidden states. While the optical probes used in this study, such as polarization rotation, ellipticity change and circular dichroism, are sufficient to characterize and verify the new non-equilibrium state with a net magnetization, further characterizations via time-resolved X-ray diffraction or X-ray magnetic circular dichroism \cite{Higley2016FemtosecondLaser, Bostedt2016LinacYears, Zayko2021UltrafastDynamics} would provide valuable details of the new magnetic order. Moreover, experimental confirmation of the phonon mode involved in inducing the new order, as well as validation of an equilibrium state with two degenerate minima, as obtained within our Ginzburg-Landau theory, would be useful to corroborate our understanding. Lastly, as the lifetime of the state exceeds 2 ms, conventional transport probes such as Hall resistivity can also be employed, thereby establishing its potential for future spintronic applications.


\section*{Methods}

\subsection*{THz pump and magneto-optical probe experiments} 
Output of Ti-sapphire amplifier (35 fs pulse duration, 800 nm wavelength, 1 kHz repetition rate) is split into pump and probe arms. Pump arm seeds an optical parametric amplifier (OPA), and the signal output of OPA at 1300 nm wavelength is used to generate THz radiation via optical rectification in an organic crystal. A set of parabolic mirrors are used to guide the THz beam and focus tightly on a diffraction limited spot of size 180$\pm$10 $\mu$m full width at half maximum (FWHM) on the sample surface at normal incidence. The THz beam spot size is measured using a THz camera (Rigi S2, Swiss Terahertz). An optical chopper was placed in the OPA beam path to measure polarization rotation $\Delta\theta$ and ellipticity change $\Delta\eta$ of the probe induced by pump (THz) pulse. The spot size of the 800 nm probe beam is 54$\pm$2 $\mu$m, measured using a CMOS camera at the sample location. The entire path of THz beam propagation and the organic crystal are kept in a box purged with dry air to avoid water absorption of THz. A translation stage placed on the probe arm (800 nm) is used to control the time delay between THz and probe pulses. For polarization rotation and ellipticity change measurements a half wave-plate (HWP) and a quarter wave-plate (QWP) were used, respectively. A Wollaston prism is placed on the path of the transmitted beam after the HWP/QWP, and the two split beams with orthogonal polarizations were detected with a pair of silicon photodetectors. For circular dichroism (CD) measurements in transmission a QWP was placed on the probe path before the sample, and the transmitted beam was detected directly by a single photodetector. The purity of the probe polarization state, both linear and circular, were carefully analysed at the sample position by using a polarizer and a powermeter. The cryostat window was placed on the path during the polarization characterization process to take into account the effect of the window on the polarization. The detailed schematics of the experimental setup is shown in Extended Data Fig.~\ref{fig:expsetupscheme}.  

\subsection*{Sample preparation}
We synthesized our FePS$_3$ crystals using a chemical vapor transport method. All the powdered elements (purchased from Sigma-Aldrich): iron (99.99\% purity), phosphorus (99.99\%) and sulfur (99.998\%), were prepared inside an argon-filled glove box. After weighing the starting materials in the correct stoichiometric ratio, we added an additional 5 wt of sulfur to compensate for its high vapor pressure. After the synthesis, we carried out the chemical analysis of the single-crystal samples using a COXEM-EM30 scanning electron microscope equipped with a Bruker QUANTAX 70 energy dispersive X-ray system to confirm the correct stoichiometry. We also checked the XRD using a commercial diffractometer (Rigaku Miniflex II). We cleaved samples before placing them into high vacuum ($\sim10^{-7}$ torr) to expose a fresh surface without contamination and oxidation. The sample thicknesses used in these experiments ranged from 20 $\mu$m to 60 $\mu$m, as determined using a profilometer.

\subsection*{Driving with off-resonant THz pulse}
To understand the role of resonant mode driving in inducing the metastable state, we performed an additional experiment with THz pulses not resonant with any of the modes in the system. For this purpose a strong THz pulse is generated by tilted-pulse front technique in LiNbO$_3$ \cite{Hebling2008GenerationPossibilities} and the polarization rotation is measured in the same way as described above. The details of the experimental setup has been reported elsewhere \cite{Zhang2023DiscoveryPerovskite}. The  spectrum of the generated pulse is given in Extended Data Fig.~\ref{fig:LNOdata}a (blue). As can be seen in Extended Data Fig.~\ref{fig:LNOdata}b, the long-lived signal is not observed when the driving field is off-resonant. This observation rules out the possibility of carrier generation upon pumping with THz as discussed in the text. 

\subsection*{Extraction of relaxation time of the metastable state}
THz pump and $\Delta \eta(T)$ probe signal is measured as a difference between the probe ellipticity in pump-on and pump-off regions. As illustrated in Fig. \ref{fig:dutycycle}c, if the relaxation time of the metastable state exceeds one millisecond, the pump-off state (white region in Fig. \ref{fig:dutycycle}c) will carry the tail of pump-on region signal, therefore giving rise to a pre-time zero difference between pump-on and pump-off regions. 

The connection between the relaxation time ($\tau_{\textrm{decay}}$) and the pre-time zero signal can be modeled as follows. We assume each pump pulse will induce a change in ellipticity signal that decays in time as $A e^{-\frac{t}{\tau_{\textrm{decay}}}}$, where $t$ is the time duration after the pump. Then the ellipticity signal before time zero for the pump-on region is given by the sum of repetitive pulses as $\sum_{n=1}^{\infty}{A e^{-\frac{n\tau_{\mathrm{pump}}}{\tau_{\textrm{decay}}}}}$, where $\tau_\mathrm{pump}$ is the temporal separation between the THz pump pulses. The pump-off state signal is similarly given by $\sum_{n=0}^{\infty}{A e^{-\frac{\tau_{\mathrm{probe}}+n\tau_{\mathrm{pump}}}{\tau_{\textrm{decay}}}}}$, where $\tau_\mathrm{probe}$ is the temporal separation between the probe pulses in the pump-on and pump-off regions. The measured pre-time zero signal is then given by: 
\begin{equation}
    \operatorname{pre-time\ zero}\ \Delta\eta = A \frac{e^{-\frac{\tau_\mathrm{pump}}{\tau_{\textrm{decay}}}}-e^{-\frac{\tau_\mathrm{probe}}{\tau_{\textrm{decay}}}}}{1-e^{-\frac{\tau_\mathrm{pump}}{\tau_{\textrm{decay}}}}}.
\label{eq:pre}
\end{equation}
In Fig.~\ref{fig:dutycycle}c, we set $\tau_\mathrm{pump}=5\ \mathrm{ms}$ and $\tau_\mathrm{probe}=1,2,3,4\ \mathrm{ms}$. The pre-time zero signal as a function of $\tau_\mathrm{probe}$ are fitted by Eq.~\ref{eq:pre} in Fig.~\ref{fig:dutycycle}d which yields $\tau_\mathrm{decay}=2.5\ \mathrm{ms}$. The ratio between the pre-time zero signal and the maximal signal after time zero is 
\begin{equation}
    \frac{\operatorname{pre-time\ zero}\ \Delta\eta}{\operatorname{maximal}\ \Delta\eta} = \frac{e^{-\frac{\tau_\mathrm{pump}}{\tau_{\textrm{decay}}}}-e^{-\frac{\tau_\mathrm{probe}}{\tau_{\textrm{decay}}}}}{1-e^{-\frac{\tau_\mathrm{probe}}{\tau_{\textrm{decay}}}}}.
\label{eq:preratio}
\end{equation}
$\tau_\mathrm{decay}$ as a function of temperature can then be obtained by numerically solving Eq.~\ref{eq:preratio}. The results for the case with $\tau_\mathrm{pump}=2\ \mathrm{ms}$ and $\tau_\mathrm{probe}=1\ \mathrm{ms}$  are shown in Fig.~\ref{fig:dutycycle}b.

\subsection*{Fitting and extracting the critical constants of the magnetization, rise time and the decay time}
For the temperature dependence of the magnetization amplitude we fit with a power law as:
\begin{align}\label{eq:Mfits}
    M(T) = A|T-T_{\textrm{N}}|^{-\gamma/2}+\textrm{const.}
\end{align}
were $A$, $\gamma$ and "$\textrm{const.}$" are free parameters. We assume $M(T)$ to be proportional to $\Delta \eta_0(T)$. $\Delta$CD$_0 (T)$ data can also be equivalently used for the fits, however we worked with ellipticity change signal instead, as it has more data points and is intrinsically less noisy, owing to its differential type detection scheme. We fit the $T>T_{\textrm{N}}$ and $T<T_{\textrm{N}}$ regions separately, and the obtained critical constants are $\gamma/2=0.53\pm0.05$ and $\gamma/2=0.56\pm0.05$, respectively (see Fig.~\ref{fig:faradayCD}e). These are in good agreement with the expected critical exponent of 3D Ising model.

Next we extract the rise time constants from the same set of data by fitting the time traces with an exponential function:
\begin{align}\label{eq:expfunc}
    \Delta \eta(t, T) = A \textrm{exp}(-\frac{t}{\tau_{\mathrm{rise}}})+B
\end{align}
the initial few picosecond window of the trace with a spike due to THz Kerr effect is omitted. Further we fit the obtained $\tau_{rise}(T)$ separately high and low temperature sides of $T_{\textrm{N}}$ with a power law given as:
\begin{align}\label{eq:tauFit}
    \tau_{\mathrm{rise}}(T) = \tau_0 \left|\frac{T_{\textrm{N}}}{T-T_{\textrm{N}}}\right|^{\nu z} + \mathrm{const.}
\end{align}
The obtained critical exponents for rise time are $\nu z$ = 1.07 $\pm$ 0.27 and $\nu z$ = 0.44 $\pm$ 0.07, for $T<T_{\textrm{N}}$ and $T>T_{\textrm{N}}$, respectively (see Fig.~\ref{fig:faradayCD}i). We fit the temperature dependence of the decay time with the same power law (Eq.\ref{eq:tauFit}), and the exponents are $\nu z$ = 0.72 and $\nu z$ = 0.56, for low and high temperature sides (see Fig.~\ref{fig:dutycycle}b). The lower temperature side critical behavior is close to the expected value from 3D Ising model ($\nu z$ = 1.27). 

\subsection*{First principles calculations}
To obtain the spin-phonon coupling of bulk FePS$_3$, we performed first principles simulations with the {\sc abinit} electronic structure code~\cite{Gonze2020TheDevelopments,Gonze1997First-principlesAlgorithm, Amadon2008Plane-waveOrbitals, Torrent2008ImplementationPressure}. We used the local density approximation with projector augmented wave (PAW) pseudopotentials, a plane wave cut-off of $20$ Ha and $40$ Ha respectively for the plane wave and PAW part, and included an empirical Hubbard $U$ of $2.7$ eV on the Fe $d$-orbitals as self-consistently determined in the {\sc octopus} electronic structure code via the ACBN0 functional. A $\Gamma$-centered Monkhorst-Pack grid with dimensions $8\times 6 \times 8$ was used to sample the Brillouin zone.

The ground state was found to have zig-zag antiferromagnetic order with spins aligned along the $z$-axis. The DFT wave functions were mapped onto effective localized orbitals via the {\sc wannier90} code, and the spin parameters were calculated by the {\sc python} package {\sc tb2j} employing the magnetic force theorem~\cite{Pizzi2020Wannier90Applications,He2021TB2J:Parameters}. The resulting spin parameters are in good agreement with recent neutron scattering data \cite{Wildes2012TheFePS3}.

Phonon frequencies and eigenvectors were calculated with {\sc abinit} for a ferromagnetic interlayer coupling, after relaxing the atomic positions and stresses to below $10^{-6}$ Ha/Bohr. We identify four relevant phonon modes at $\Omega_1 = 2.67$ THz, $\Omega_2 = 3.27$ THz, $\Omega_3 = 4.77$ THz and $\Omega_4 = 7.48$ THz, in excellent agreement with our experimental THz and Raman scattering data (see also Extended Data Table~\ref{tab:phonon_parameters}). The spin-phonon couplings for these four modes were calculated in the frozen phonon approximation and are given in Extended Data Table~\ref{tab:spin_phonon}. These large spin-phonon couplings  are consistent with previous works on magnon-phonon hybridization \cite{liuPRL2021, vaclavkova2021, Zhang2021CoherentInsulator}, and they likely arise from the same origin as the large single-ion anisotropy, i.e. the unquenched Fe orbital magnetic moments as reported in Ref.~\cite{Lee2023GiantFePS3}.

\subsection*{Spin Monte Carlo simulations}
To validate the exchange parameters obtained from first principles calculations, as well as the effective spin-phonon Hamiltonian, we performed simulated annealing using the Monte Carlo Metropolis algorithm. The calculations were done for a monolayer or thin bulk (up to $10$ layers) of FePS$_3$ using the equilibrium spin parameters of Extended Data Table~\ref{tab:spin_parameters} and a spin length $S = 2$, for in-plane supercells of linear sizes between $L = 10$ or $20$. The simulated annealing was initialized at a temperature $T \approx 170$ K and performed down to a target temperature $T_0 = 70$ K in steps of $\Delta T \approx 2$ K. At each temperature we performed $2000$ Monte Carlo sweeps to thermalize the system, followed by $4000$ measurements performed at an interval of $50$ sweeps to obtained thermodynamic averages. Our calculations find a magnetic transition from the paramagnetic to the zigzag antiferromagnetic state at $T_\textrm{N} \approx 115$ K, in good agreement with experiment, thereby validating the calculated equilibrium spin parameters.

To assess the effects of a spin-phonon coupling we modified the magnetic parameters according to the spatial modulation patterns shown in Fig.~\ref{fig:criticality}b and Extended Data Fig.~\ref{fig:otherphononfluence}. Such a simulations corresponds to a frozen phonon approximation, and reveals the expected magnetic properties for a fixed phonon displacement $Q$. A more detailed discussion of the simulations is provided in Supplementary Note 6.

\subsection*{Multiscale modeling of dynamics}
The time scales involved in our experiments span the range from typical magnetic scales of ps to relaxation dynamics happening over several ms. As there is no single method that can access all these timescales, we have used a multi-scale approach where different methods are combined to cover different regimes. Our microscopic methods (spin Monte Carlo and Landau-Lifshitz dynamics) are restricted to $\sim 1$ ns, while solutions to the stochastic Ginzburg-Landau equations can be obtained up to $\sim 10$ ns. These methods (see Supplementary Notes 9 and 10) were used to address the short to intermediate time scales, covering the initial excitation and the build-up of the THz-induced magnetization. We note that all these theories agree in the prediction that a finite magnetization appears if and only if the phonon mode $Q_2$ is displaced away from equilibrium, either in a static (as corroborated by thermal Monte Carlo simulations) or in dynamic fashion (as corroborated by Landau-Lifshitz and Ginzburg-Landau dynamics). To understand the nature and microscopic dynamics of the magnetized state up to time scales of $\sim 10$ ns, these methods are therefore both accurate and sufficient.

To address the long-time dynamics, we combine results from the theory of dynamic critical phenomena with analytical and numerical solutions of the (stochastic) Ginzburg-Landau equations. These methods are expected to provide an accurate description of the relaxation dynamics, which is dominated by the interplay of long-wavelength excitations and thermal fluctuations and therefore insensitive to microscopic details. Specifically, as all our theories find the dominant antiferromagnetic order parameter $L$ to be largely unaffected by the sub-dominant ferromagnetic order parameter $M$ and the phonon displacement $Q_2$, the critical dynamics of $L$ (in the region $T \sim T_{\rm N}$) is determined solely by its intrinsic dynamics. We can therefore employ the standard theory of dynamical critical phenomena~\cite{Hohenberg1977}, which predicts that perturbations of the dominant order parameter decay as $\delta L(t) \sim \exp(-t/\tau)$ with $\tau \sim |T-T_{\rm N}|^{-\nu z}$.

As discussed in the next section, we find that the coupling term $gLMQ_2$ leads to a critical slowing down also in the dynamics of $M$. This is supported by numerical solutions of the stochastic Ginzburg-Landau equation for $M$ (for times up to $\sim 10$ ns), as discussed in Supplementary Note 10. Therefore, for times $t \gtrsim 1$ ns but $t < \tau$, we have $M(t) \sim \exp(-t/\tau)$ with the same lifetime as $L$.

\subsection*{Critical slowing down of magnetization dynamics}
An intuitive way to understand the critical slowing down of $M$ is to note that for $T \to T_N$, the fast order parameter ($M$) quickly decays to the local minimum defined by the slow order parameter ($L$). Once $M$ has reached this minimum, it adiabatically follows it back to equilibrium, such that $M(t) \approx \sqrt{(g^2 L(t)^2/\Omega - a)/b}$ (see Supplementary Note 10). Here we have suppressed the subscripts on $a_M$ and $b_M$. Since the relaxation of the free energy minimum is determined by the dynamics of the antiferromagnetic order, it exhibits a critical slowing down. More precisely, for slow variations of $L$ close to criticality, we can approximately solve the deterministic Ginzburg-Landau equation for $M$ by replacing $a$ with $a - g^2 L(t)^2/\Omega$. Assuming that $L(t) = L(0) e^{-t/\tau}$, as is well known from the theory of dynamical critical phenomena~\cite{Hohenberg1977}, and defining $\lambda^2 = g^2 L(0)^2/\Omega$, the solution is then
\begin{align}
 M(t) &\approx \sqrt{\frac{\lambda^2 e^{-t/\tau} - a}{b}} \bigg[1 - e^{2at - 2\lambda^2 \tau e^{-t/\tau}} \bigg( 1 + \frac{\lambda^2 e^{-t/\tau} - a}{bM(0)^2} \bigg) \bigg]^{-1/2}
\end{align}
The leading order term of this equation recovers the simple adiabatic approximation discussed above, while the sub-leading terms show that there is a crossover at $at \approx \lambda^2 \tau e^{-t/\tau}$ to a regime of intrinsic exponential decay $M(t) \sim e^{-at}$. Importantly, close to the critical temperature, $M(t)$ will decay with with a diverging relaxation time $\tau \sim |T - T_N|^{-\nu z}$. Therefore, for times $t \gtrsim 1$ ns but $t < \tau$, we have $M(t) \sim \exp(-t/\tau)$ with the same lifetime as $L$. In Supplementary Note 10 we provide further evidence of the critical slowing down of the magnetization, and how it is inherited from the critical dynamics of $L$, based on numerical solutions of the stochastic Ginzburg-Landau equations.
\\

\section*{Data Availability}
The datasets generated and/or analysed during the current study are available from the corresponding author on reasonable request

\bibliography{feps3}

\newpage
\section*{Acknowledgments}
We thank Alfred Zong, Bryan Fichera, Dominik Juraschek, Honglie Ning, Martin Eckstein and Zhanybek Alpichshev for fruitful discussions. We acknowledge the support from the US Department of Energy, Materials Science and  Engineering  Division,  Office  of  Basic  Energy  Sciences  (BES  DMSE)  (data taking and analysis), and Gordon and Betty Moore Foundation’s EPiQS Initiative grant GBMF9459 (instrumentation and manuscript writing).  E.V.B. acknowledges funding from the European Union's Horizon Europe research and innovation programme under the Marie Skłodowska-Curie grant agreement No 101106809. A.R. was supported by the Cluster of Excellence Advanced Imaging of Matter (AIM), Grupos Consolidados (IT1249-19), SFB925, “Light Induced Dynamics and Control of Correlated Quantum Systems,” and the Max Planck Institute New York City Center for Non-Equilibrium Quantum Phenomena. Z. Z. and K. A. N. acknowledge support from the U.S. Department of Energy, Office of Basic Energy Sciences, under Award No. DESC0019126. The work at SNU was supported by the Leading Researcher Program of Korea’s National Research Foundation (Grant No. 2020R1A3B2079375). 

\section*{Author Contributions}
B.I., T.L. and N.G. conceived the study. B.I. and T.L. designed and built the experimental setup and performed the measurements. E.V.B. and A.R. performed the first principle calculations and Monte Carlo simulations. Z.Z. performed  experiments with THz generated from LiNbO$_3$ under the supervision of K.A.N. J.P. and J.K. synthesized and characterized FePS$_3$ single crystals under the supervision of J.-G.P. B.I., T.L. and A.v.H performed the data analysis. B.I., T.L., A.v.H and N.G. interpret the data and wrote the manuscript with critical inputs from E.V.B. and A.R. and all other authors. The project was supervised by N.G.   

\section*{Competing Interests}
The authors declare no competing interests.

\newpage

\begin{center}
\begin{figure}
   \sbox0{\includegraphics{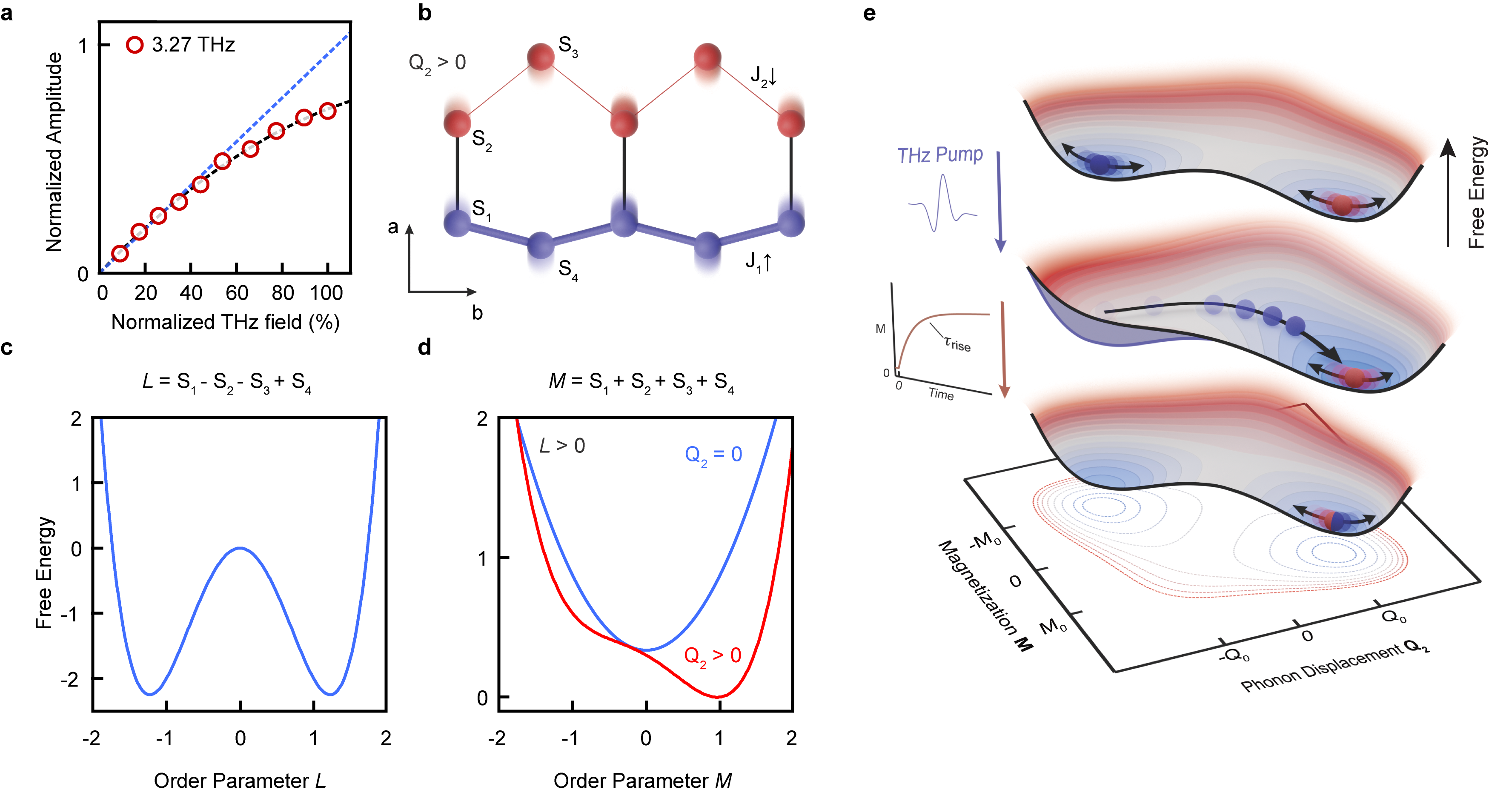}}
    \begin{minipage}{\wd0}
  \usebox0
  \captionsetup{labelfont={bf},labelformat={default},labelsep=period,name={Fig.}}
  \caption{\textbf{Nonlinear excitation of $Q_2$ phonon triggers a net magnetization and the critical fluctuations facilitate its metastability.}  Field strength dependence of the 3.27 THz (\textbf{a}) phonon mode amplitude. The 3.27 THz phonon has both linear and quadratic components, that interfere destructively, which is manifested in the "curved down" shape of the field dependence. \textbf{b.} Real-space motions of the $\mathrm{Fe^{2+}}$ ions due to excitation of the $Q_2$ phonon and the spatial pattern of the phonon-modulated exchange couplings. An enhancement of the magnetic interaction on a given bond is illustrated by a thick line and up-pointing arrow, whereas a decrease is shown by a thin line a down-pointing arrow. Labels $S_i$ and $J_i$ denote the relevant spin and exchange parameters of the magnetic unit cell. \textbf{c} and \textbf{d} are the magnetic free energy landscapes as a function of zig-zag AFM order parameter \textit{L} and magnetization \textit{M}, respectively. For a fixed 3.27 THz phonon displacement $Q_2$, the system develops to degenerate energy minima at finite $L$ and $M$, such that the sign of $M$ is determined by the signs of $Q_2$ and $L$  through a term $Q_2LM$ in the free energy. \textbf{e.} Equilibrium free energy landscape obtained from effective Ginzburg-Landau theory. In thermal equilibrium the order parameter fluctuates between the two degenerate minima, thereby preventing the build-up of a finite magnetization (upper panel). When driven into an asymmetric configuration by distorting the lattice (middle panel), a single minimum is favored leading to a net magnetization (lower panel). The subsequent dynamics are governed by the critical fluctuations, which results in a critically slowed down relaxation back to equilibrium. 
  \label{fig:criticality}}
\end{minipage}
\end{figure}  
\end{center}

\newpage
\FloatBarrier
\begin{center}
\begin{table}[ht]
 \begin{tabularx}{\columnwidth/2}{m|o|o|o|o}
   \hline Mode & $Q_1$ & $Q_2$ & $Q_3$ & $Q_4$ \\ \hline
  $\Omega^{(c)}$ (THz) & 2.67 & 3.27 & 4.77 & 7.48 \\ \hline
  $\Omega^{(c)}$ (meV) & 11.1 & 13.5 & 19.8 & 31.0 \\ \hline
  Irrep. & $B_g$ & $A_g$ & $A_g$ & $B_g$  \\ \hline
 \end{tabularx}
 \begin{minipage}{\columnwidth/2}
  \usebox0
 \captionsetup{labelfont={bf},labelformat={default},labelsep=period,name={Extended Data Table}}
 \caption{{\bf Phonon properties of FePS$_3$.} Calculated mode frequencies and irreducible representations of the Raman active phonon modes of FePS$_3$, with $C_{2h}$ point group. Phonons proximate to the lower magnon branch at $3.69$ THz are listed.}
 \label{tab:phonon_parameters}
 \end{minipage}
\end{table}
\end{center}
\FloatBarrier

\FloatBarrier
\begin{center}
\begin{table}[ht]
 \begin{tabularx}{\columnwidth/2}{m|n|n|n|n} \hline\hline
  & \multicolumn{4}{c}{Equilibrium exchange coupling (meV)} \\ \hline
  & $J_1$ & $J_2$ & $J_3$ & $J_4$ \\ \hline
  & -0.81 & -1.38 & -0.81 & -1.38 \\ \hline\hline
  \multicolumn{4}{c}{} & \\ \hline\hline
  & \multicolumn{4}{c}{Spin-phonon coupling (meV/{\AA})} \\ \hline
  Mode & $\alpha_1$ & $\alpha_2$ & $\alpha_3$ & $\alpha_4$ \\ \hline
  $Q_1$ &  -5.4 &  -5.0 &   5.7 &   3.7 \\ \hline
  $Q_2$ &  17.1 &   0.5 & -17.6 &  -0.4 \\ \hline
  $Q_3$ &   9.7 & -20.7 &   9.7 & -20.8 \\ \hline
  $Q_4$ & -30.0 & -29.0 &  26.1 &  26.4 \\ \hline\hline
 \end{tabularx} 
 \begin{minipage}{\columnwidth/2}
  \usebox0
 \captionsetup{labelfont={bf},labelformat={default},labelsep=period,name={Extended Data Table}}
 \caption{{\bf Nearest neighbor exchange interactions and spin-phonon couplings in FePS$_3$.} Calculated magnetic exchange interactions of the four inequivalent nearest neighbor bonds indicated in Fig.~\ref{fig:criticality}b, as well as the spin-phonon coupling coefficients $\alpha_i$ labeled in correspondence with $J_i$.}
 \label{tab:spin_phonon}
 \end{minipage}
\end{table}
\end{center}
\FloatBarrier

\FloatBarrier
\begin{table}
 \begin{tabularx}{\columnwidth/2}{o|o|o|o|o|o}
 \hline\hline
  $J_1^F$ & $J_1^A$ & $J_2^F$ & $J_2^A$ & $J_3$ & $\Delta$ \\ \hline
  -0.56 & -1.40 & 0.01 & 0.43 & 0.53 & 2.46\\
  \hline\hline
 \end{tabularx}
 \begin{minipage}{\columnwidth/2}
  \usebox0
  \captionsetup{labelfont={bf},labelformat={default},labelsep=period,name={Extended Data Table}}
 \caption{{\bf Equilibrium magnetic interaction parameters of FePS$_3$.} The magnetic exchange interactions for nearest neighbor (subscript $1$), next nearest neighbor (subscript $2$) and third nearest neighbor interactions (subscript $3$) used to obtain the equilibrium properties of FePS$_3$ via simulated annealing. The superscript denotes the interactions on intra-chain (ferromagnetic, $F$) and inter-chain (antiferromagnetic, $A$) bonds. The single-ion anisotropy is denoted by $\Delta$.}
 \label{tab:spin_parameters}
 \end{minipage}
\end{table}
\FloatBarrier

\begin{center}
\begin{figure}[!htbp]
   \setcounter{figure}{3}
   \sbox0{\includegraphics{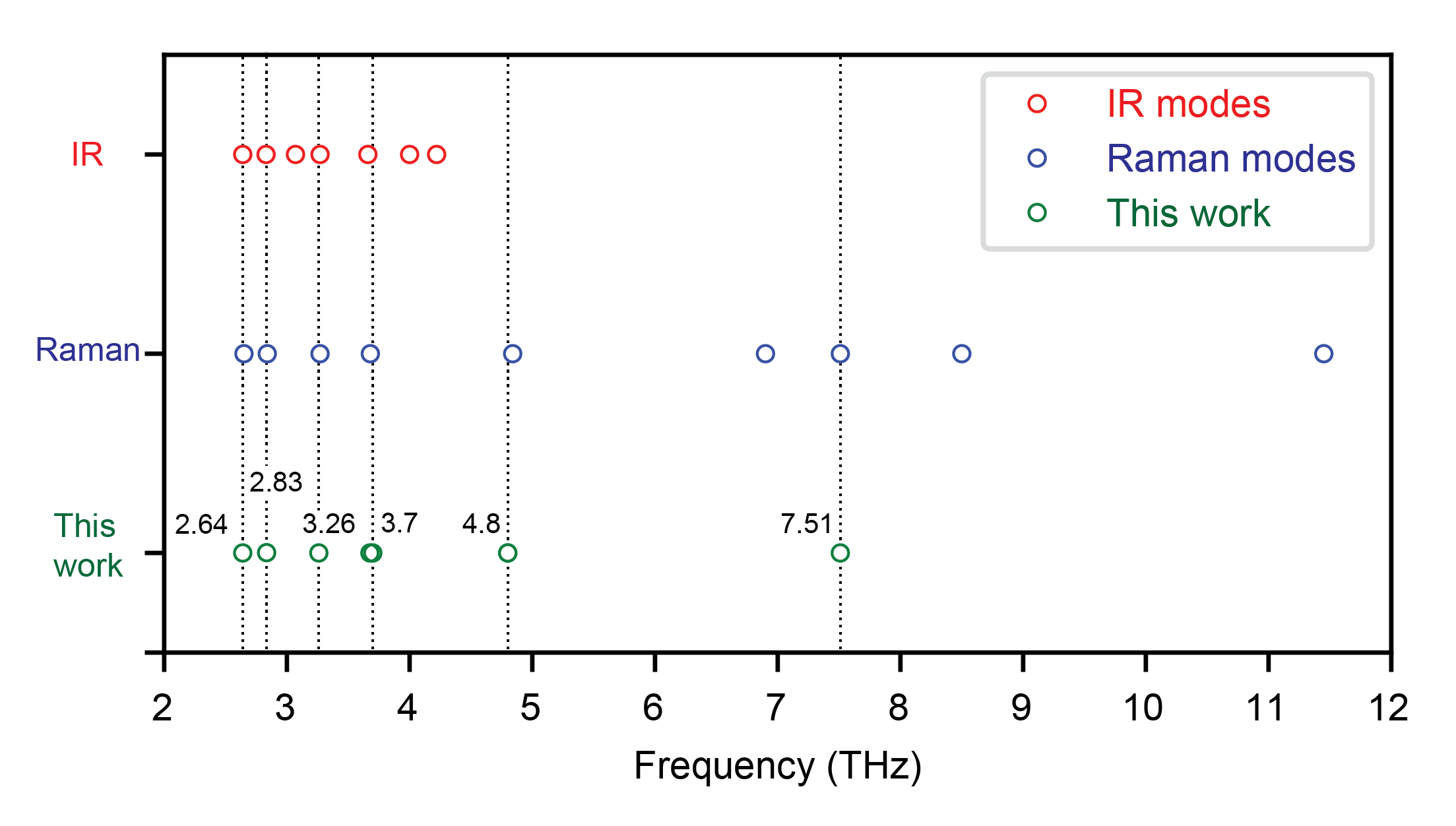}}
    \begin{minipage}{\wd0}
  \usebox0
  \captionsetup{labelfont={bf},labelformat={default},labelsep=period,name={Extended Data Table }}
  \caption{\textbf{List of low energy modes.} Mode frequencies as detected by Raman \cite{Lee2016Ising-TypeFePS3} and infrared \cite{vaclavkova2021, Zhang2021CoherentInsulator} spectroscopies. Out of seven modes observed in this work, 4.8 THz and 7.51 THz modes are only observed in Raman spectra, whereas the other three phonon modes (2.64 THz, 2.83 THz, and 3.27 THz) and two magnon modes near 3.69 THz appear in both Raman and infrared measurements. See Extended Data Figure 9. Error bars are smaller than the marker sizes.
  \label{fig:RamanIRTable}}
\end{minipage}
\end{figure}  
\end{center}

\newpage
\begin{center}
\begin{figure}[!htbp]
   \setcounter{figure}{0}
   \sbox0{\includegraphics{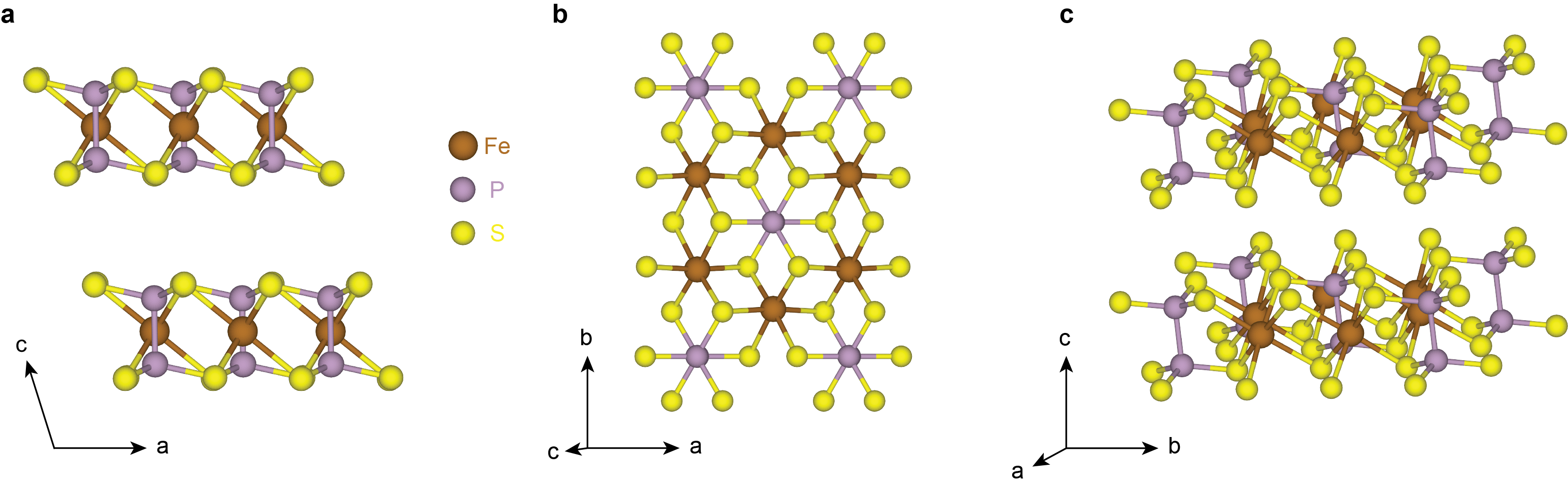}}
    \begin{minipage}{\wd0}
  \usebox0
  \captionsetup{labelfont={bf},labelformat={default},labelsep=period,name={Extended Data Fig.}}
  \caption{\textbf{Crystal structure of FePS$_3$.} \textbf{a.} Crystal structure schematics of the \textit{a-c} plane projected along \textbf{b$^{\ast}$}-axis. \textbf{b.} \textit{a-b} plane projected along \textbf{c$^{\ast}$}-axis. \textbf{c.} 3D view of the crystal structure. All the structures were generated by VESTA software \cite{Momma2011VESTAData}.
  \label{fig:crystalStructure}} 
\end{minipage}
\end{figure}  
\end{center}

\begin{center}
\begin{figure}[!htbp]
   \setcounter{figure}{1}
   \sbox0{\includegraphics{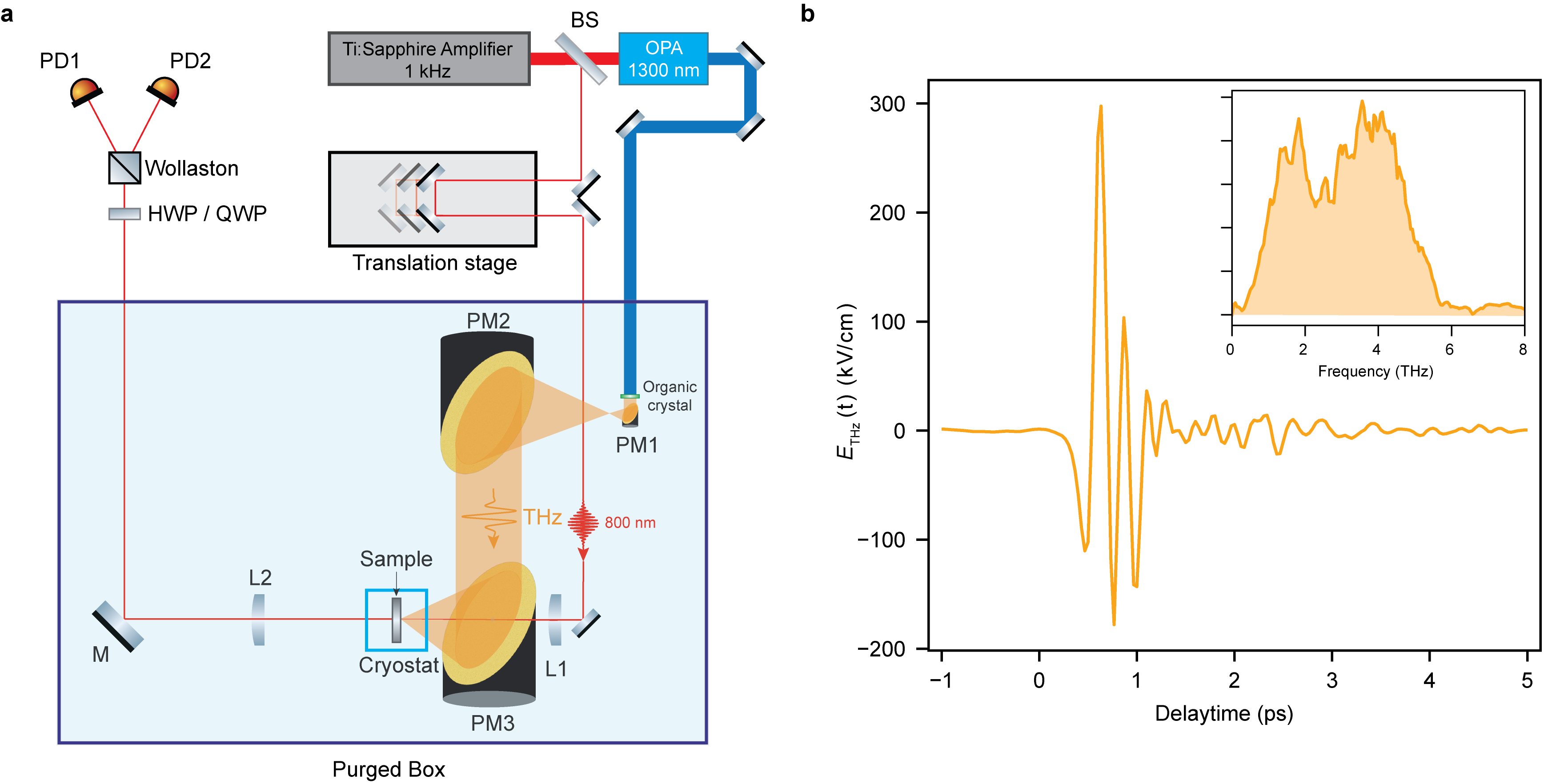}}
    \begin{minipage}{\wd0}
  \usebox0
  \captionsetup{labelfont={bf},labelformat={default},labelsep=period,name={Extended Data Fig.}}
  \caption{\textbf{Experimental setup.} \textbf{a.} Detailed schematics of the experimental setup. The abbreviations used here: BS - beam splitter; OPA - optical parametric amplifier; PM - parabolic mirror; L - lens; HWP - half wave plate; QWP - quarter wave plate; PD - photo-diode.  \textbf{b.} Field profile of the THz pulse. Peak field strength is $\sim$300 kV/cm. Fourier spectrum of the field is shown in the inset.
  \label{fig:expsetupscheme}} 
\end{minipage}
\end{figure}  
\end{center}

\begin{center}
\begin{figure}[!htbp]
   \setcounter{figure}{2}
   \sbox0{\includegraphics{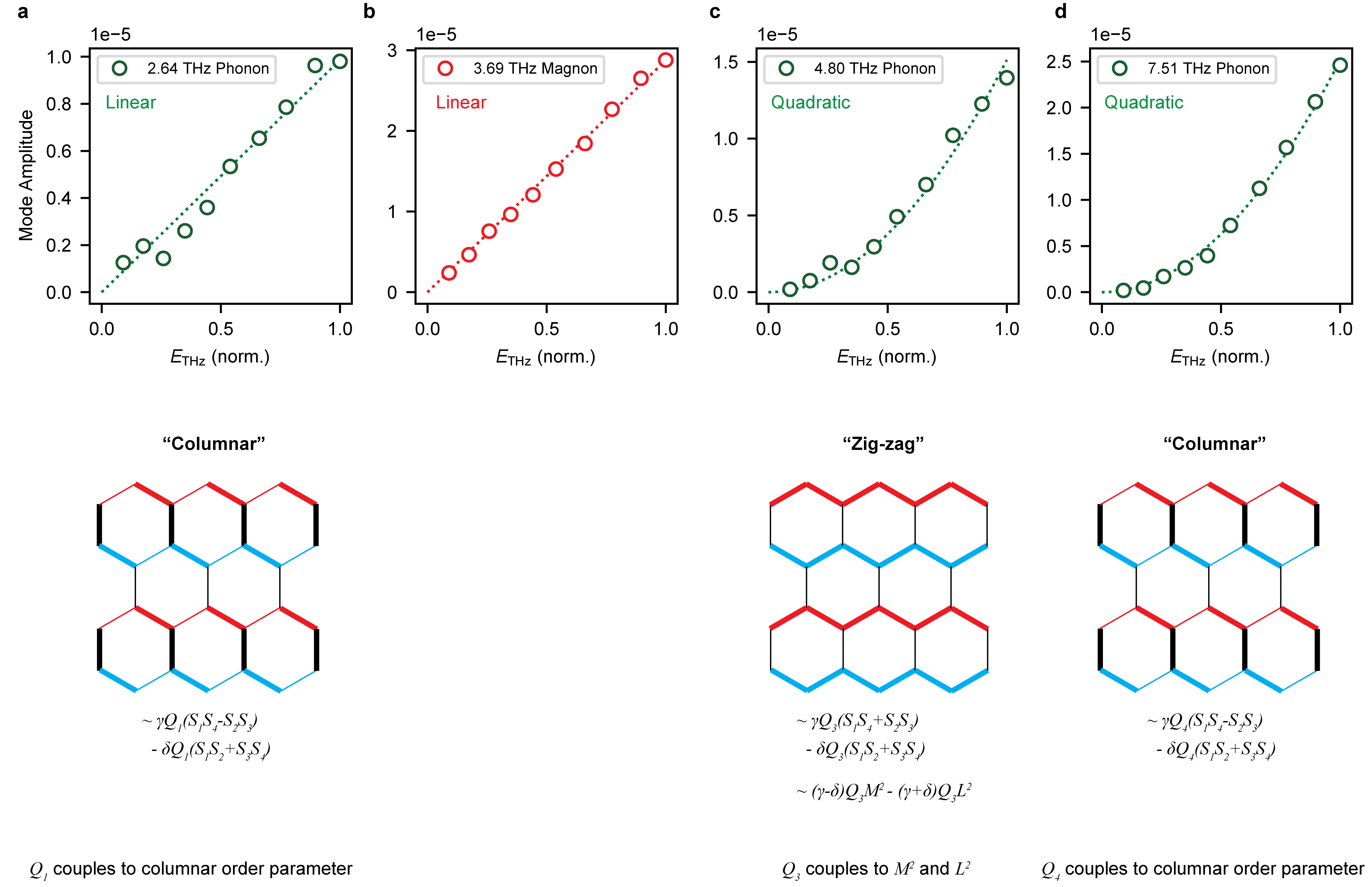}}
    \begin{minipage}{\wd0}
  \usebox0
  \captionsetup{labelfont={bf},labelformat={default},labelsep=period,name={Extended Data Fig.}}
  \caption{\textbf{Field strength dependencies of other low energy modes.} \textbf{a}, \textbf{b}, \textbf{c}, and \textbf{d} are THz field strength dependencies of the 2.64 THz phonon, 3.69 THz magnon, 4.80 THz phonon, and 7.51 THz phonon mode amplitudes, respectively. Dashed lines in \textbf{a} and \textbf{b} are linear fits, whereas in \textbf{c} and \textbf{d} the fits are quadratic. The lower panels are the modulation patterns of the nearest-neighbour exchange interactions under the corresponding phonon displacements. The $Q_1$ (2.64 THz) and $Q_4$ (7.51 THz) phonons induce so called columnar modulations (see Supplementary Note 5). Since the columnar phase is far from the zig-zag phase, the contributions of these two phonons to the Ginzburg-Landau free energy can be neglected. The $Q_3$ (4.80 THz) phonon has a "zig-zag" type modulations, and it couples individually to $M^2$ and $L^2$. Unless there is a mechanism that breaks the degeneracy between opposite signs of magnetization, the $Q_3$ is also not expected to generate a net magnetization. These findings are further supported by spin Monte Carlo simulations (see Supplementary Note 6), and therefore, we neglect their contributions in the main text.
  \label{fig:otherphononfluence}} 
\end{minipage}
\end{figure}  
\end{center}

\begin{center}
\begin{figure}[!htbp]
   \setcounter{figure}{3}
   \sbox0{\includegraphics{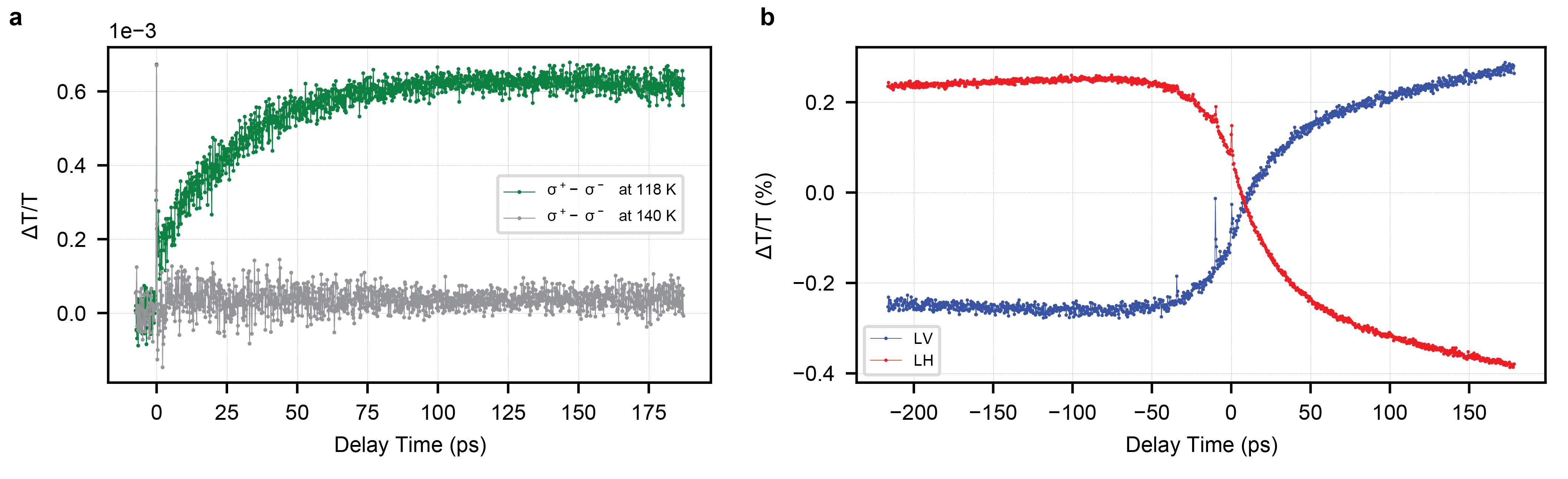}}
    \begin{minipage}{\wd0}
  \usebox0
  \captionsetup{labelfont={bf},labelformat={default},labelsep=period,name={Extended Data Fig.}}
  \caption{\textbf{THz field-induced circular and linear dichroism.} \textbf{a}. THz-induced circular dichroism $\Delta$CD measurements at 118K (green) and at 140K (grey). $\Delta$CD is measured as a difference in transmittance of right-circularly polarized (RCP) and left-circularly polarized (LCP) probe beams. At high temperature (140K) $\Delta$CD signal is absent, indicating there are no experimental artifacts in our setup that would result in $\Delta$CD signal. \textbf{b}. Suppression of equilibrium linear dichroism by THz pulse at 118K. In equilibrium, transmission of horizontally polarized light is 2.3 times larger than the vertically polarized light, which reduces by $\sim$1\% after THz pumping, as shown in \textbf{b}. Since LD is proportional to the square of zigzag AFM order parameter \cite{Zhang2021ObservationFePS3}, we can conclude that the THz pulse suppresses the AFM order by about $\sim$0.5\%.
  \label{fig:CDLD}}
\end{minipage}
\end{figure}  
\end{center}


\begin{center}
\begin{figure}[!htbp]
   \setcounter{figure}{4}
   \sbox0{\includegraphics{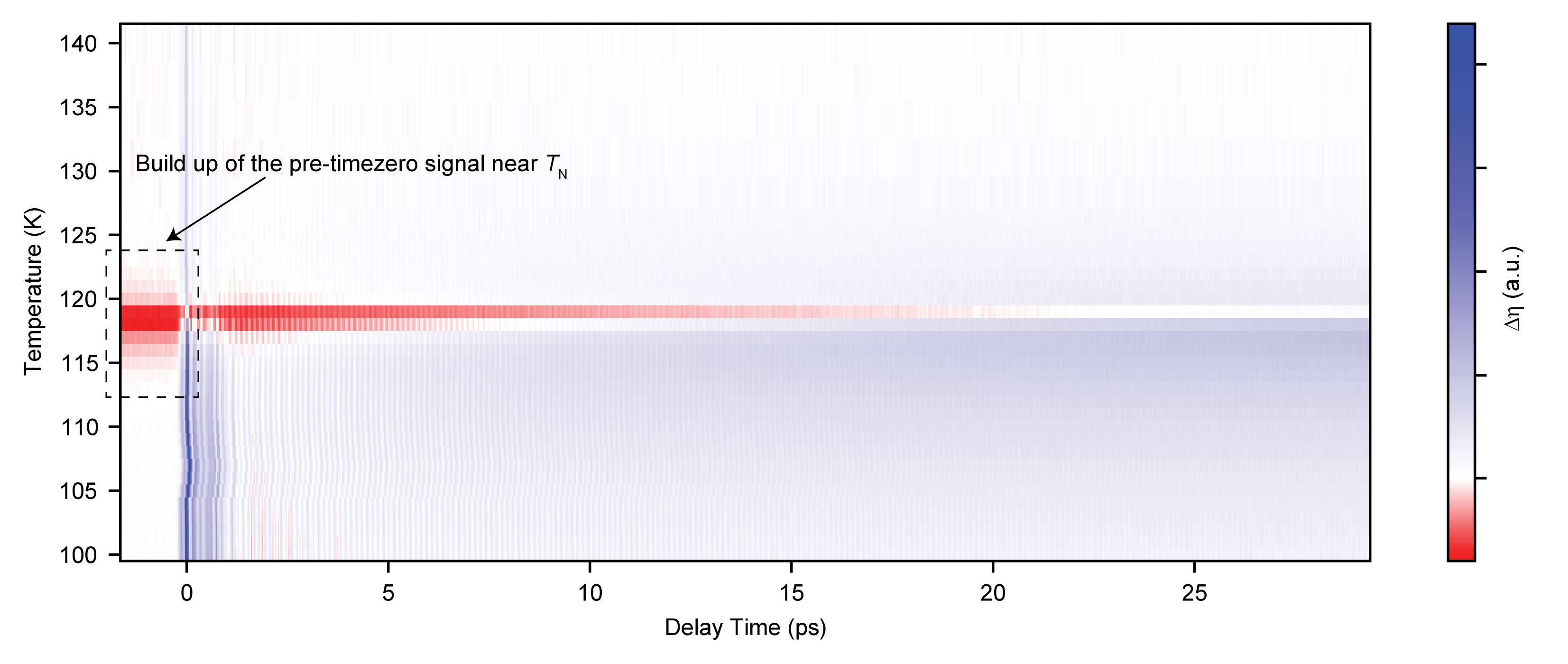}}
    \begin{minipage}{\wd0}
  \usebox0
  \captionsetup{labelfont={bf},labelformat={default},labelsep=period,name={Extended Data Fig.}}
  \caption{\textbf{Temperature dependence of ellipticity change time traces.} Raw data of temperature dependent $\Delta\eta$ experiments performed near N\'eel temperature with fine steps. It shows a marked change near transition point. A box with black dashed lines marks the region near transition a build-up of pre-time zero signal. Fig.~\ref{fig:faradayCD}e in the Main Text is obtained by measuring the signal amplitude at long time delays ($\sim$170 ps). The pre-timezero signal decreases with temperature for $T<118K$, whereas it increases with increasing temperature for higher temperatures.}
  \label{fig:rawtimetrace}
\end{minipage}
\end{figure}  
\end{center}


\begin{center}
\begin{figure}[!htbp]
   \setcounter{figure}{5}
   \sbox0{\includegraphics{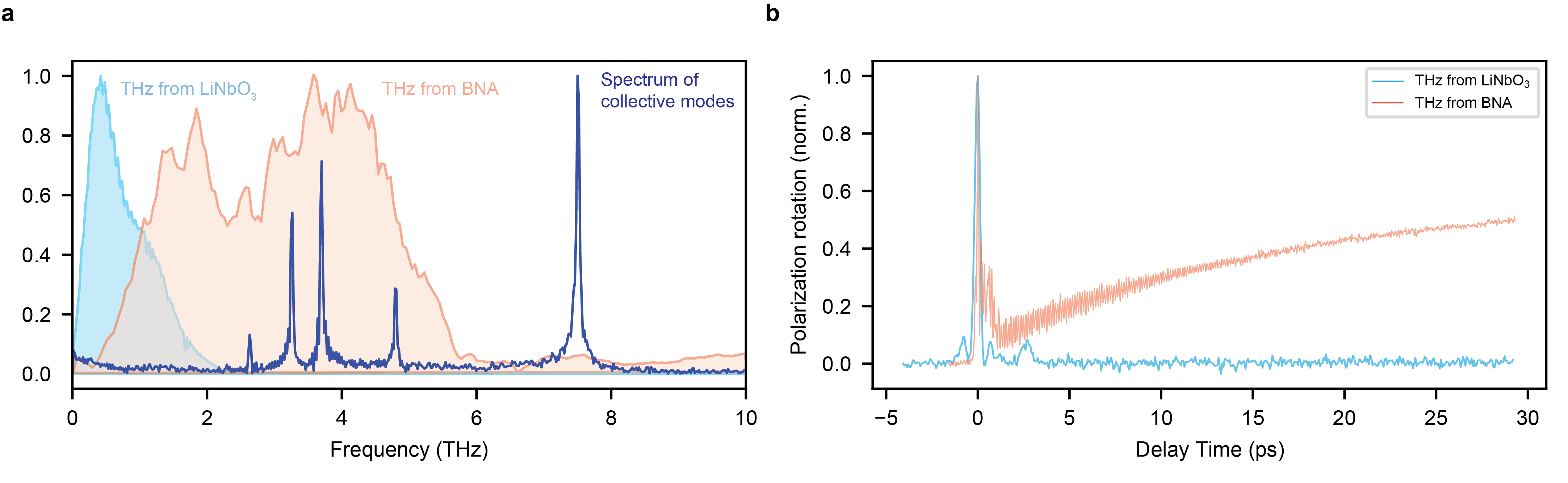}}
    \begin{minipage}{\wd0}
  \usebox0
  \captionsetup{labelfont={bf},labelformat={default},labelsep=period,name={Extended Data Fig.}}
  \caption{\textbf{Importance of nonlinearly excited low energy modes.} We repeat the same experiment with single-cycle THz excitation pulses generated from LiNbO$_3$ by tilted-pulse-front technique \cite{Hebling2008GenerationPossibilities}. The THz spectrum is given in \textbf{a} (blue), that is below the lowest phonon observed in FePS$_3$. In \textbf{b} we compare signals induced by two different THz pulses generated by organic crystal (BNA) and LiNbO$_3$ (see Methods) at $T = $118 K, and in the case of LiNbO$_3$ we did not observe any long-lived signal.
  \label{fig:LNOdata}}
\end{minipage}
\end{figure}  
\end{center}

\newpage
\begin{center}
\begin{figure}
   \setcounter{figure}{6}
   \sbox0{\includegraphics{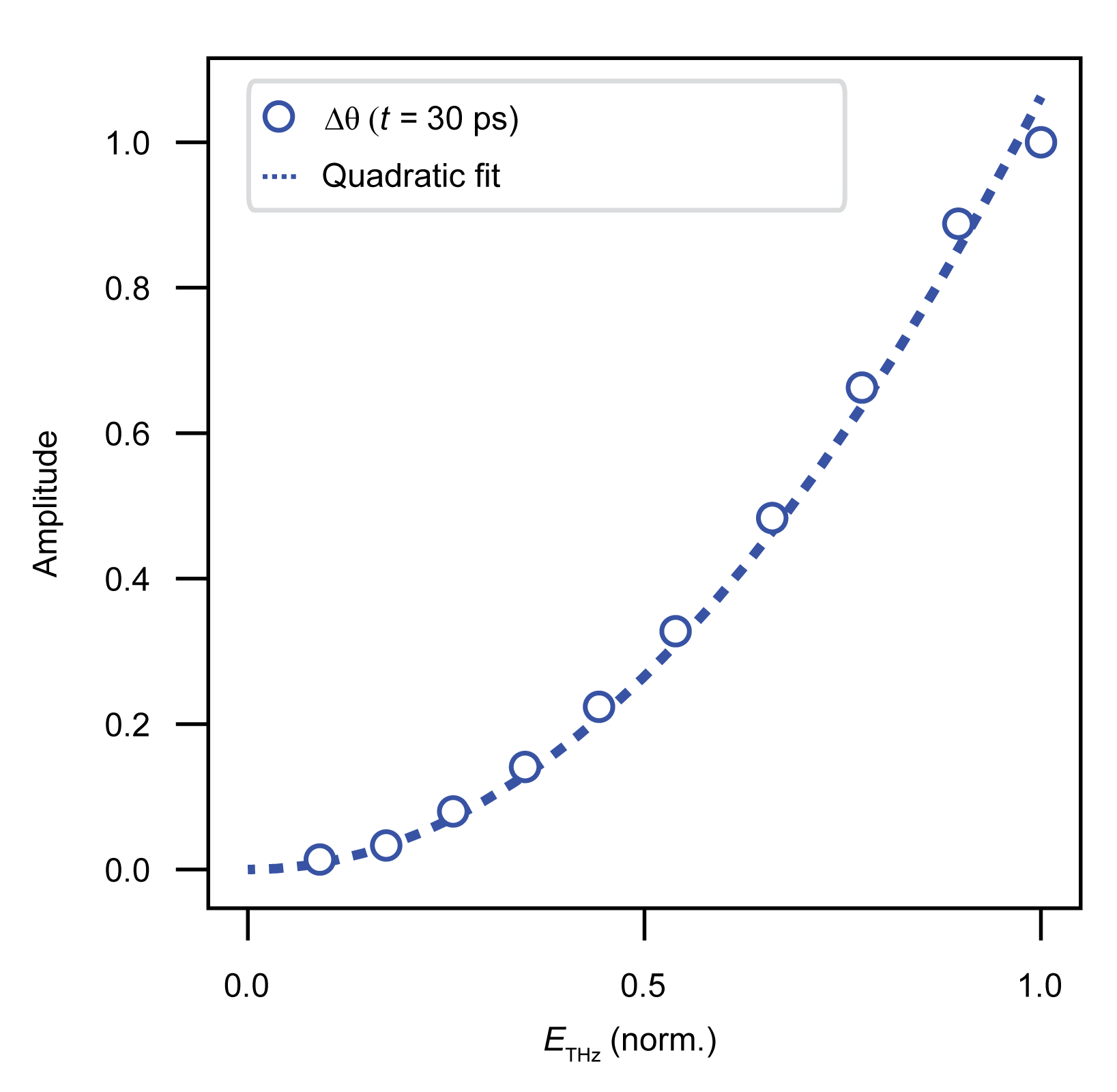}}
    \begin{minipage}{\wd0}
  \usebox0
  \captionsetup{labelfont={bf},labelformat={default},labelsep=period,name={Extended Data Fig.}}
  \caption{\textbf{THz field strength dependence of the metastable signal.} The amplitude of the metastable signal is quadratic in THz field amplitude. This is in agreement with the expected linear dependence of the induced $M$ on the phonon displacement $Q_2$, whereas $Q_2$ has an excitation pathway quadratic on $E_\textrm{THz}$. Hence $M$ is expected to scale quadratically with $E_\textrm{THz}$.
  \label{fig:signalfluence}}
\end{minipage}
\end{figure}  
\end{center}

\newpage
\begin{center}
\begin{figure}
   \setcounter{figure}{7}
   \sbox0{\includegraphics{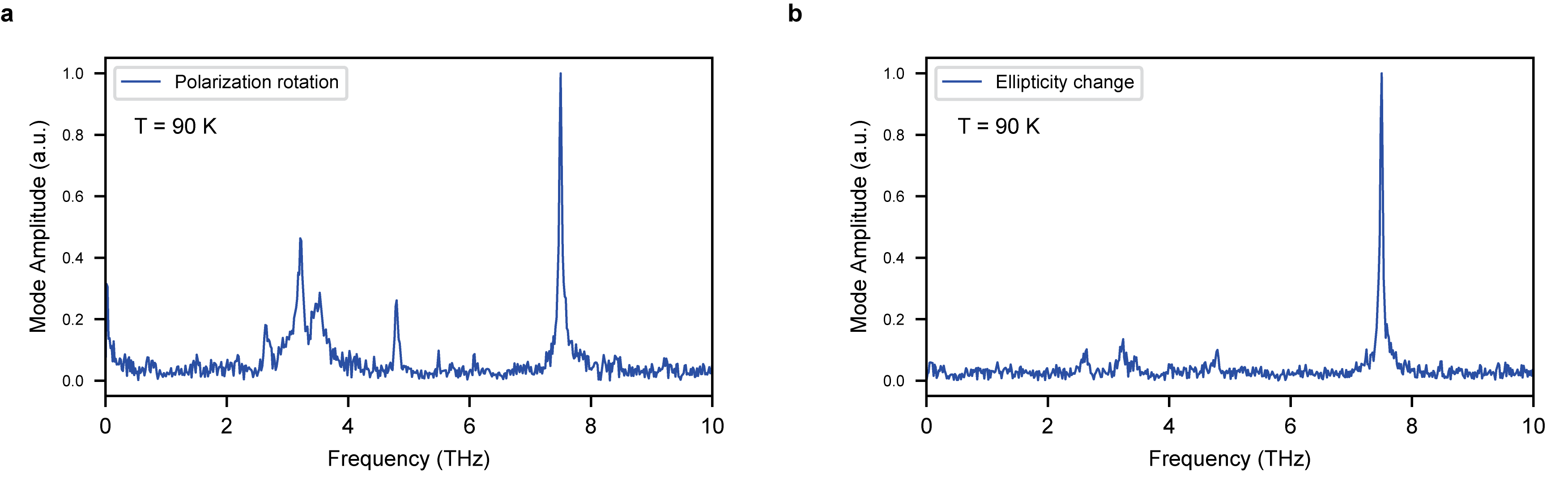}}
    \begin{minipage}{\wd0}
  \usebox0
  \captionsetup{labelfont={bf},labelformat={default},labelsep=period,name={Extended Data Fig.}}
  \caption{\textbf{Mode spectra measured in the polarization rotation and ellipticity change channel at 90K.} Spectra of collective excitations as measured in the polarization rotation (\textbf{a}) and ellipticity change (\textbf{b}) channels at 90K. The detection scheme in these two channels use half waveplate (HWP) and quarter waveplate (QWP), respectively. The relative amplitudes of the modes in the energy window between 2 THz to 5 THz, are more pronounced in the polarization rotation channel.
  \label{fig:modesHWPQWP}}
\end{minipage}
\end{figure}  
\end{center}

\newpage
\begin{center}
\begin{figure}
   \setcounter{figure}{8}
   \sbox0{\includegraphics{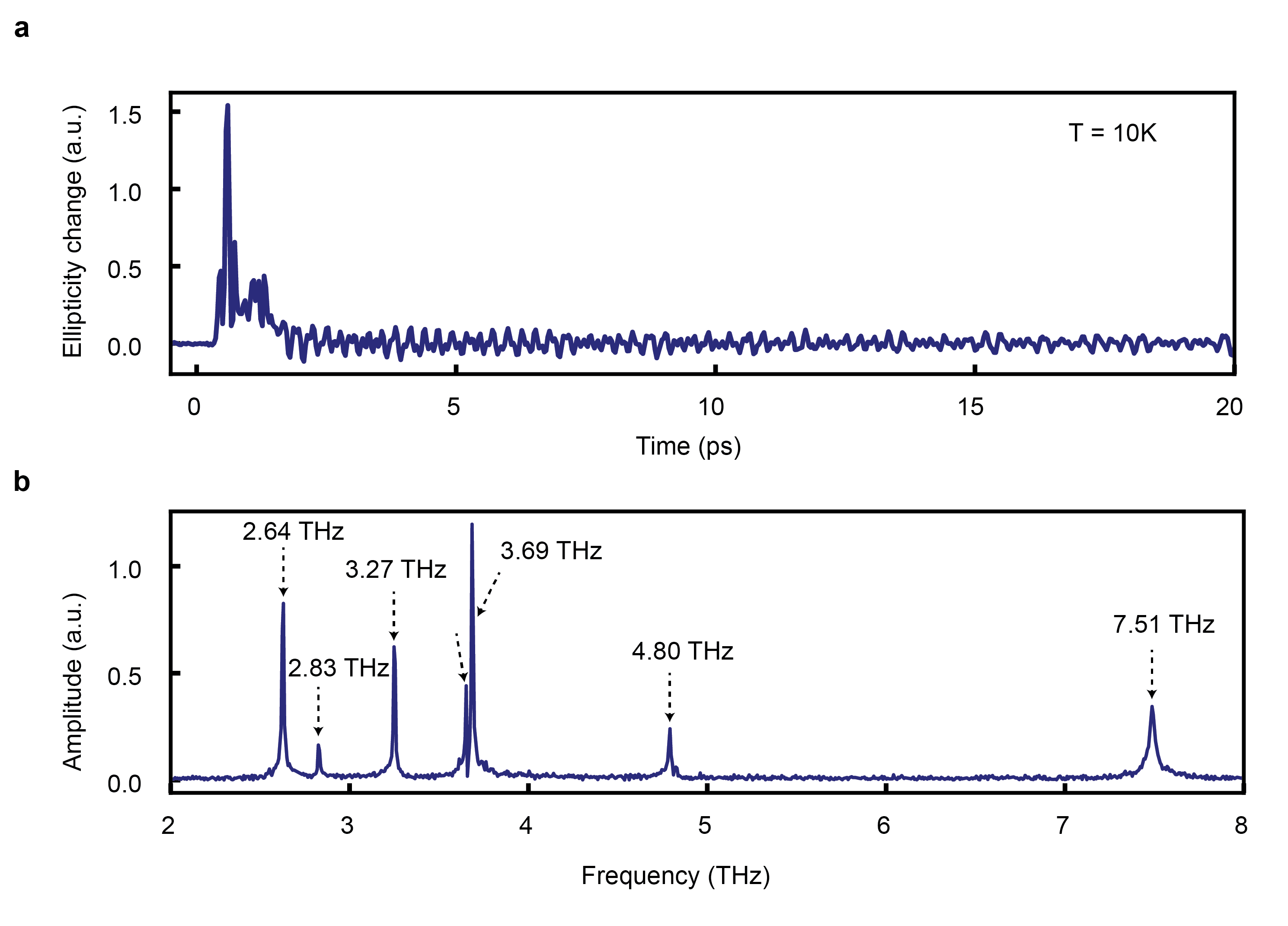}}
    \begin{minipage}{\wd0}
  \usebox0
  \captionsetup{labelfont={bf},labelformat={default},labelsep=period,name={Extended Data Fig.}}
  \caption{\textbf{Mode spectra measured in the ellipticity change channel at 10K with wider temporal window.} \textbf{a.} A time trace of THz-induced ellipticity change signal with a longer temporal window. \textbf{b.} Fourier transform of the oscillations in \textbf{a}. Owing to the longer time window, the frequency resolution is enhanced, which enabled to resolve the splitting of the two magnon energies by $\sim$0.033 THz = 0.13 meV. Additionally, in this set of measurements, we can observe the mode at 2.83 THz.
  \label{fig:fullspectrum}}
\end{minipage}
\end{figure}  
\end{center}

\end{document}